\documentclass[aps,twocolumn,prx,floatfix]{revtex4-2}

\usepackage{xcolor,hyperref}
\hypersetup{
   colorlinks,
   linkcolor={blue!50!black},%{red!80!black},
   citecolor={blue!50!black},
   urlcolor={blue!80!black}
}
\usepackage{epsfig,amsmath,amssymb}
\usepackage{bm}
\usepackage{soul,ulem}
\usepackage{hyperref}
\usepackage{footmisc}
\usepackage{wrapfig}
\DeclareMathAlphabet{\mathitbf}{OML}{cmm}{b}{it}
\newcommand{\pa}{\partial}

\renewcommand{\=}{\!=\!}
\DeclareMathAlphabet\mathbfcal{OMS}{cmsy}{b}{n}
\begin{document}

\title{Local micromechanics in a mean-field model of glasses\\ reveal key properties of its non-equilibrium RSB phase}
\author{Makoto Suda$^1$}
\author{Edan Lerner$^2$}
\author{Eran Bouchbinder$^3$}
\email{eran.bouchbinder@weizmann.ac.il}
\affiliation{$^1$Department of Earth Science, Tohoku University, Sendai, Miyagi 980-8578, Japan\\
$^2$Institute for Theoretical Physics, University of Amsterdam, Science Park 904, 1098 XH Amsterdam, the Netherlands\\
$^3$Chemical and Biological Physics Department, Weizmann Institute of Science, Rehovot 7610001, Israel}

\begin{abstract}
A recently formulated mean-field model of glasses features an equilibrium, zero-temperature Replica-Symmetry-Breaking (RSB) transition in some parameter range. In this range, the model's solution in the Replica-Symmetric phase is expressed in terms of an effective, self-consistent random potential for uncoupled degree of freedoms, where the transition to the RSB phase is characterized by the emergence of spectral-edge localized modes and a pseudogapped quartic vibrational spectrum, resulting in a finite spin-glass susceptibility. These properties are preserved in numerical solutions of the model under non-equilibrium conditions, i.e., upon an instantaneous quench. Inspired by recent advances in computer glasses, we define a micromechanical response function --- the linear response to local force monopoles --- in the framework of the mean-field model. We establish exact relations between the force monopole stiffness and global susceptibilities, which suggest a close correspondence between the non-equilibrium RSB phase of the model and the above-mentioned effective random potential description. As such, the obtained micromechanical observables constitute a concrete realization of the collective degrees of freedom of the model, offering a bridge between a glassy mean-field model and finite-dimensional glasses. We show that the model's vibrational spectrum can be computed solely from the monopole response statistics and, by employing a marginal stability criterion, we extract a characteristic frequency/stiffness scale of soft glassy modes, which is related to the boson peak in finite-dimensional, laboratory glasses.
\end{abstract}

\maketitle

\section{B\lowercase{ackground and motivation}}
\label{sec:Intro}

Glasses constitute a distinct phase of matter generically formed by rapidly cooling high-temperature liquids~\cite{Goldstein1969-ao,Debenedetti2001-an,Cavagna2009-wq,Dyre2026-lp}. The resulting materials are out-of-equilibrium, i.e., their properties depend on their formation history, and solid-like, i.e., they can sustain shear stresses over enormous timescales, but feature no long-range order. The intrinsically disordered and non-equilibrium nature of glasses poses serious challenges. While some continuum-scale glass properties related to global symmetries --- such as long-wavelength phononic modes, elasticity and sound attenuation~\cite{Schirmacher1998-fa,Schirmacher2007-ya,Marruzzo2013-zu,DeGiuli2014-oi,Mizuno2018-jy,scattering_jcp,Cui2020-bx,jcp_letter_scattering_2021,Baggioli2022-oo,Vogel2023-rv,Vogel2026-xe} --- may be understood using perturbative approaches, many other aspects of glass physics depend on smaller-scale disorder and various emergent localization phenomena~\cite{lerner2026nonlinear} that are not yet fully understood.

It is now established that the low-frequency vibrational spectra of glasses host, in addition to spatially-extended phonons that exist due to global translational symmetries, also nonphononic, quasi-localized modes~\cite{Lerner2016-ra,Mizuno2017-sc,Angelani2018-vt,Wang2019-jq,Bonfanti2020-nv,Lerner2021-en,Wang2025-so}. These low-frequency glassy modes follow a non-Debye vibrational density of states (VDoS) that features a universal $\sim\!\omega^4$ tail for small vibrational frequencies $\omega$ and a peak at a higher frequency. It is also established that plasticity in glasses is mediated by localized irreversible rearrangements~\cite{Kobayashi1980-uu,Falk1998-fm,Malandro1999-zm,Maloney2006-um,Schuh2007-ti,Falk2010-ax,Hentschel2010-aa} and that glasses driven by sufficiently strong mechanical forces undergo a nonlinear yielding transition~\cite{Utz2000-ry,Varnik2004-gv,Karmakar2010-ft,Lin2014-ov,Rainone2015-dd,Jin2018-cy,Ozawa2018-gp,Nicolas2018-xj,Berthier2025-hv}. Moreover, upon employing cyclic deformation, glasses can store mechanical memories, which can be subsequently and robustly extracted~\cite{Fiocco2014-jq,Hima-Nagamanasa2014-nt,Regev2015-vq,Adhikari2018-dl,Keim2019-ez,Adhikari2025-lb,Paulsen2025-av}. All these diverse glassy phenomena, including nonphononic vibrational spectra, yielding transitions and memory formation appear to be manifestations of an underlying complex energy landscape, with generic properties associated with glassy disorder and glassiness.

Recently, a disordered Hamiltonian mean-field model, which belongs to the class of soft spin-glass models, has been introduced~\cite{Bouchbinder2021-dh,Rainone2021-am}. The model features no translational symmetry, i.e., it a priori excludes phononic degrees of freedom, and as such, focuses on nonphononic ones. It predicted in a unified manner a broad range of  low-temperature properties of glasses, both in the absence of external driving and in its presence~\cite{Bouchbinder2021-dh,Rainone2021-am,Folena2022-tn,Moriel2024-wp,Maimbourg2024-lb,Suda2025-kl}. This progress raises the need both to obtain a deeper understanding of the model and to challenge it in relation to other aspects of glass physics.

Our goal here is two-fold; first, we aim at studying the statistical micromechanics of the mean-field model, inspired by significant recent advances in computer glasses~\cite{Lerner2018-me,Rainone2020-jo,Rainone2020-gs}. By ``micromechanics" we refer to response functions to mechanical perturbations applied on the scale of a single degree of freedom. Second, we aim at obtaining a deeper analytic understanding of the model under a broader range of equilibrium and non-equilibrium conditions.

Specifically, while an equilibrium, zero-temperature solution in the Replica-Symmetric (RS) phase --- upon approaching the Replica-Symmetry-Breaking (RSB) transition --- is available in some parameter range~\cite{Bouchbinder2021-dh,Folena2022-tn}, we study the model deep in the RSB phase following an instantaneous, non-equilibrium quench. The equilibrium, zero-temperature replica solution is expressed in terms of an effective, self-consistent random potential for uncoupled degree of freedoms, which has been shown to remain valid right beyond the RSB transition~\cite{Folena2022-tn}. In addition, some qualitative resemblance between the zero-temperature behavior of the model and its off-equilibrium dynamics following a quench has been indicated~\cite{Folena2022-tn}. We show that indeed the statistical micromechanics of the model feature mathematical similarities to the equilibrium effective potential and reveal key properties of its non-equilibrium RSB phase, as will be explained in detail below.

\section{T\lowercase{he mean-field model and its established properties}}
\label{sec:model}

To set the stage for our analysis, we start by briefly presenting the disordered Hamiltonian mean-field model of~\cite{Bouchbinder2021-dh,Rainone2021-am}, which takes the form
\begin{equation}
    H\!=\!\frac{1}{2}\sum_{i=1}^{N} k_i x_i^2 + \frac{1}{4!}\sum_{i=1}^{N} x_i^4 + \sum_{i<j}J_{ij}x_ix_j -f(t)\sum_{i=1}^{N} x_i\ .
    \label{eq:Hamiltonian}
\end{equation}
Here, the continuous displacements $\{x_i\}$ represent small groups of particles in an equilibrium liquid that oscillate at a frequency $\sqrt{k_i}$ around $x_i\=0$ in the absence of interactions. The stiffness $k_i$ follows a uniform distribution $p(k)$ over the interval $[k_{\rm m}, k_0]$, with $k_0\!>\!k_{\rm m}\!\ge\!0$. Upon quenching the liquid, long-range elasticity builds up and the coarse-grained oscillators interact. This interaction is captured at the mean-field level through the bilinear energy term $\sum_{i<j}J_{ij}x_i x_j$, where $J_{ij}$ are Gaussian, i.i.d.~random variables of zero mean and variance $J^2/N$.

The self-organization processes occurring during the quench are mimicked by minimizing $H$, which displaces the oscillators from their initial $x_i\=0$ position and gives rise to a non-trivial energy landscape. The existence/stability of the obtained minima is ensured by adding the leading-order stabilizing anharmonicity $\sum_i x_i^4$, featuring no disorder. Finally, the term proportional to $f(t)$ may account either for a constant quenched field $h$ applied prior to the minimization of $H$~\cite{Bouchbinder2021-dh,Rainone2021-am,Folena2022-tn,Moriel2024-wp,Maimbourg2024-lb} or for a time-dependent field that is applied after the minimization and represents external driving forces~\cite{Suda2025-kl}.

For a finite field $f(t)\=h\!>\!0$ and an initially gapped stiffness distribution, $k_{\rm m}\!>\!0$, the model features an RS phase for some range of $h$ and $J$ values in the equilibrium, zero-temperature limit~\cite{Bouchbinder2021-dh}. The replica method solution is then expressed in terms of decoupled oscillators in an effective, self-consistent random potential of the form~\cite{Bouchbinder2021-dh}
\begin{equation}
    v_{\rm eff}(x)=\frac{1}{2}\left(k-J^2\chi^{(\rm RS)}\right) x^2 + \frac{1}{4!}x^4 -(\delta{\!f} + h)x \ ,
    \label{eq:eff_potential}
\end{equation}
where $\delta{\!f}$ is a Gaussian random force of zero mean and a variance that is self-consistently determined~\cite{Bouchbinder2021-dh}. $J^2\chi^{(\rm RS)}$ is a stiffness down-shift due to collective interactions, and the solution is valid for $k\!-\!J^2\chi^{(\rm RS)}\!\ge\!0$, where $\chi^{(\rm RS)}(J)$ is self-consistently determined according to
\begin{equation}
\chi^{(\rm RS)}=\left\langle \frac{1}{v''_{\rm eff}[x^*(\delta{\!f},k\!-\!J^2\chi^{(\rm RS)})]}\right\rangle_{\delta{\!f},k-J^2\chi^{(\rm RS)}} \ .
\end{equation}
Here, $x^*(\delta{\!f},k\!-\!J^2\chi^{(\rm RS)})$ is the absolute minimum of $v_{\rm eff}(x)$.

The model undergoes an RSB transition upon crossing a critical line $J_{\rm c}(h)$ in the $(h,J)$ plane, reflected in the statistical properties of the minima of $H$; specifically, the eigenvalues spectrum of the Hessian $\bm{\mathcal{M}}\!\equiv\!\pa^2H/\pa\bm x\pa\bm x$ and the localization properties of the eigenmodes. The $J_{\rm c}(h)$ line is split into two qualitatively different parts separated by a special point corresponding to $k_{\rm m}-J^2\chi^{(\rm RS)}\=0$, where double-wells in the effective potential emerge, and to a vanishing replicon eigenvalue (see~\cite{Bouchbinder2021-dh} for definition). Below this point, the nonphononic VDoS ${\cal D}_{\!_{\rm G}\!}(\omega)$ features a pseudogapped quadratic tail ${\cal D}_{\!_{\rm G}\!}(\omega)\!\sim\!\omega^2$, populated by delocalized modes, similarly to other mean-field models~\cite{Crisanti1992-xn,Cavagna1998-qa,Franz2015-cc,Ikeda2022-ol}. Yet, above the special point on the $J_{\rm c}(h)$ line, a quartic ${\cal D}_{\!_{\rm G}\!}(\omega)\!\sim\!\omega^4$ tail populated by localized modes emerges, in agreement with a broad range of observations in finite-dimensional computer glasses~\cite{Lerner2016-ra,Mizuno2017-sc,Angelani2018-vt,Wang2019-jq,Bonfanti2020-nv,Lerner2021-en,Wang2025-so}.

The model under non-equilibrium conditions, i.e., upon minimizing $H$ of Eq.~\eqref{eq:Hamiltonian} without any thermalization in a way that mimics an instantaneous quench, has been extensively studied in the RSB phase using numerical simulations~\cite{Rainone2021-am,Moriel2024-wp,Suda2025-kl}. Recent theoretical analyses indicate that some properties of the zero-temperature equilibrium model are preserved under these conditions~\cite{Folena2022-tn}. Indeed, the numerical studies provided strong evidence that the quartic ${\cal D}_{\!_{\rm G}\!}(\omega)\!\sim\!\omega^4$ tail populated by localized modes persists deep inside the non-equilibrium RSB phase. In fact, these properties persist also for $k_{\rm m}\=0$, i.e., when the initial liquid-like stiffness distribution is gapless, and for $h\=0$, for which an RS phase does not exist at all. We hereafter set $h\=0$ and $k_{\rm m}\=0$, i.e., employ $p(k)\=k_0^{-1}$ in the interval $[0, k_0]$. This model features a single intrinsic parameter $J/k_0$ (hence we hereafter set $k_0\=1$), quantifying quenched disorder.

It is precisely this $h\=0$ and $k_{\rm m}\=0$ version of the model that has been shown to feature additional remarkable glassy properties, in agreement with a broad range of computational and laboratory observations, both in the non-driven and strongly driven regimes. Specifically, it has been shown that while the aforementioned model's VDoS features a universal quartic power-law ${\cal D}_{\!_{\rm G}\!}(\omega)\=A_{\rm g}(J)\,\omega^4$ in the $\omega\!\to\!0$ limit, where $A_{\rm g}(J)$ is a non-universal, disorder-dependent prefactor, it also features a generic peak at a somewhat higher frequency $\omega_{\rm p}$, which is related to the so-called ``boson peak''~\cite{Schirmacher1998-fa,Schirmacher2007-ya,DeGiuli2014-oi,Malinovsky1986-vb,Ramos1997-ew,Nakayama2002-ve,Chumakov2004-gb,Ramos2022-nf,Festi2026-zu,Marruzzo2013-zu,Gurarie2003-qi,Parisi2003-ex,Gurevich2003-oo,Parshin2007-od,Lerner2023-dp,Gonzalez-Jimenez2023-un,Moriel2024-wp}. Moreover, while eigenmodes at the $\omega\!\to\!0$ edge of spectrum feature the aforementioned strong localization manifested in $e\!\sim\!{\cal O}(1/N)$, where $e\!\equiv\!(N\sum_i\psi_i^4)^{-1}$ is the participation ratio of a normalized eigenmode ${\bm \psi}$, eigenmodes at $\omega_{\rm p}$ are delocalized, featuring $e\!\sim\!{\cal O}(N)$.

Very recently, the model has also been studied in the strongly nonlinear driven regime, specifically with a periodic/oscillatory force $f(t)$ of variable amplitude $f_0$, in the limit of a vanishing loading-rate. It has been found that the model predicts essentially all the salient features of the oscillatory yielding phase transition and of mechanical memory formation observed in computer and laboratory glasses~\cite{Suda2025-kl}. These include an absorbing-to-diffusive transition in the system's microscopic trajectories and large-scale hysteresis above a disorder-dependent critical yielding amplitude $f_0\!>\!f_{\rm y}(J)$, dynamic slowing down from both sides of the transition, mechanical annealing effects, and memories storage and their robust extraction~\cite{Fiocco2014-jq,Hima-Nagamanasa2014-nt,Regev2015-vq,Adhikari2018-dl,Keim2019-ez,Adhikari2025-lb,Regev2013-ff,Fiocco2013-ak,Kawasaki2016-nb,Leishangthem2017-sw,Yeh2020-ve,Bhaumik2021-mr,Liu2022-hq,Parley2022-id,Parley2024-bz,Cochran2024-ms,Jocteur2024-ke,Paulsen2025-av}.

In view of the above, we aim here at addressing several important questions. First, in light of the success of the model in predicting a broad range of glassy phenomena, we aim at further testing its predictive powers in relation to its statistical micromechanical properties. This effort will be made in the context of significant recent advances in understanding the statistical micromechanics of finite-dimensional computer glasses~\cite{Lerner2018-me,Rainone2020-jo,Rainone2020-gs}, to be discussed below. We will also aim at making additional analytic progress in understanding the model, to be demonstrated in the derivation and validation of a mean-field identity that relates local micromechanical properties to global susceptibilities of different orders.

Second, we are interested in understanding whether some aspects of the mathematical structure and properties of the random potential in Eq.~\eqref{eq:eff_potential}, in light of some recent suggestions~\cite{Folena2022-tn}, are nevertheless preserved in the latter phase. Moreover, if the latter is indeed the case, one is interested in identifying the actual objects, or observables in the model, which correspond to the effective degrees of freedom in the random potential. As we will show below, quite surprisingly, all these questions are interlinked.

\section{T\lowercase{he force monopole response}}
\label{sec:force_monople}

As explained above, many aspects of glass physics involve disorder-induced localization and emergent mesoscopic lengthscales. A particular challenge in this context is to disentangle these physical effects from those related to long-wavelength phonons. In recent years, various mesoscopic disorder quantifiers and micromechanical response quantities have been developed and applied to computer glasses, leading to new insights and understanding~\cite{Lerner2018-me,Rainone2020-jo,Rainone2020-gs,Tanguy2010-at,Manning2011-tu,Ding2014-tr,Patinet2016-ya,Zylberg2017-ey,Richard2020-vr,Kapteijns2021-ec,Kapteijns2021-gv,Giannini2021-ll,Richard2023-bo,Gonzalez-Lopez2023-jx,lerner2026nonlinear}. One of these is the glass linear response to local dipolar perturbations~\cite{Lerner2018-me,Rainone2020-jo,Rainone2020-gs}. That is, consider any pair $\alpha$ of interacting particles in a zero-temperature glass, i.e., in one of its inherent structures, and apply to them a force dipole ${\bm d}_{\alpha}$ (a procedure sometimes termed ``pinching'' the glass~\cite{Rainone2020-jo}). The linear response to this force dipole is given by ${\bm z}_{\alpha}\=\bm{\mathcal{M}}^{-1}\!\cdot{\bm d}_{\alpha}$, where $\bm{\mathcal{M}}$ is the glass Hessian. As force dipoles are net-zero (gradient-like) sources, they naturally suppress rigid-body motions and hence screen long-wavelength elastic modes. As such, their linear response provides access to mesocopic elastic properties of glasses~\cite{Lerner2018-me,Rainone2020-jo,Rainone2020-gs}.

\begin{figure*}[t]
    \centering
    \includegraphics[width=0.8\paperwidth]{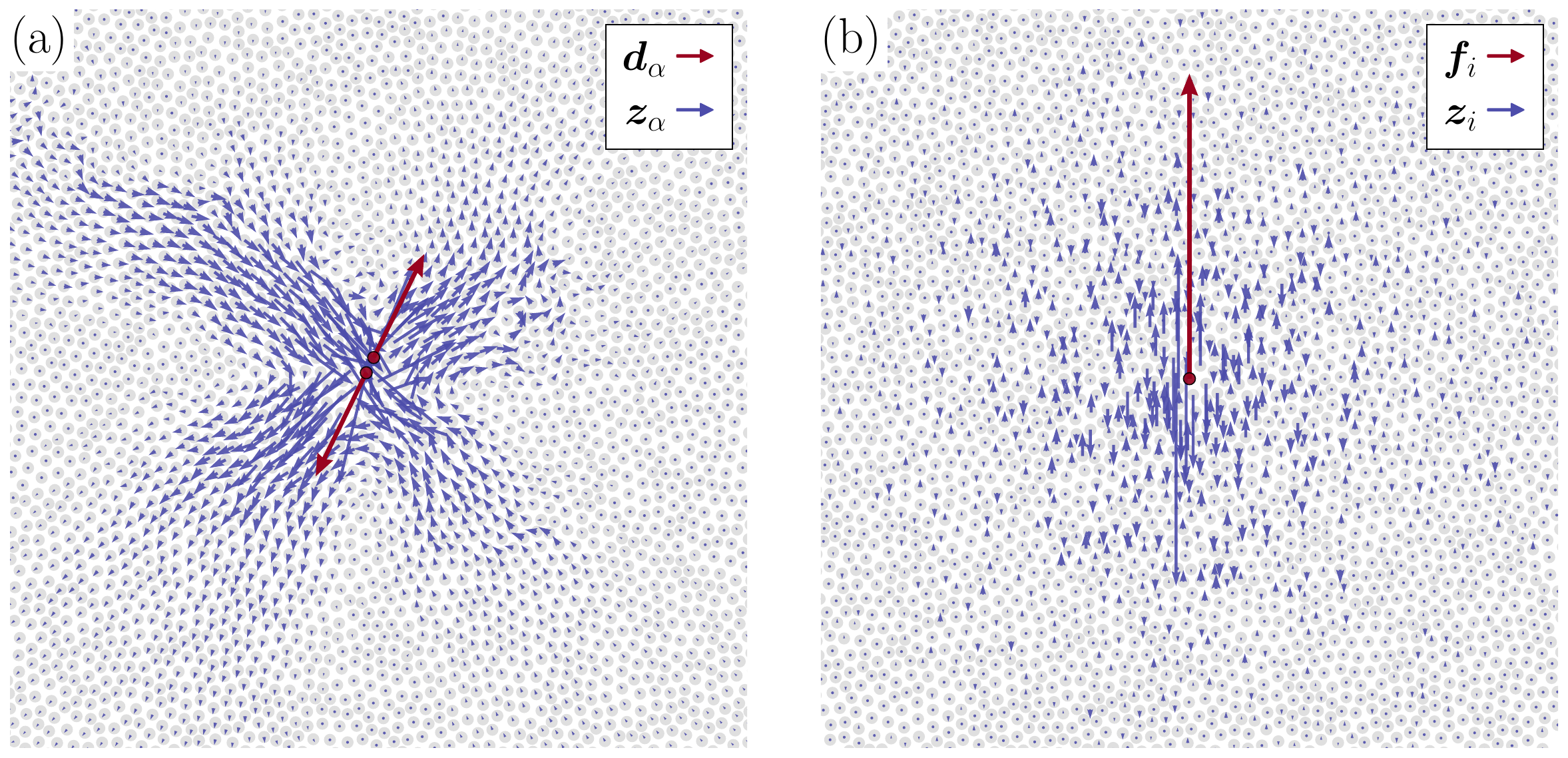}
    \caption{(a) The displacement response $\bm z_\alpha$ (blue arrows) to a local force dipole $\bm d_\alpha$ (red arrows) in a 2D computer glass with spatially-decaying, finite-range interactions~\cite{Lerner2018-me}, see text. The response is localized near the force dipole. (b) A 2D representation of the displacement response $\bm z_i$ (blue arrows) to a local force monopole ${\bm f}_i$ (red arrow) in the mean-field model, where the $i^{\tiny \rm th}$ oscillator is placed at the center and the response arrows are oriented relative to the imposed force. While the mean-field model features no spatial structure, and specifically it is characterized by statistically uniform all-to-all interactions, by arranging the the remaining oscillators such that those with small $\kappa_{j \neq i}$ are placed closer to the force, a localized structure is revealed. It is localized in the conventional sense that only a rather small fraction of the oscillators carry the response (note that only half of the $N\!=\!4096$ is shown), analogously to the dipole response ${\bm z}_\alpha$ shown in panel (a). This indicated localization is quantified in Fig.~\ref{fig_monopole_short:Fig2} using the participation number.}
    \label{fig_monopole_short:Fig1}
\end{figure*}

Indeed, as shown in Fig.~\ref{fig_monopole_short:Fig1}a for a 2D computer glass, and as has been extensively discussed in~\cite{Lerner2018-me,Rainone2020-jo,Rainone2020-gs}, the linear response field ${\bm z}_{\alpha}$ features a non-trivial spatial structure that closely resembles nonphononic, low-frequency quasi-localized modes. The stored energy associated with ${\bm z}_{\alpha}$ takes the form $\tfrac{1}{2}\,\kappa_{\rm dipole}^{(\alpha)}\,|\bm z_{\alpha}|^2$, where $\kappa_{\rm dipole}^{(\alpha)}\=({\bm z}_{\alpha}\cdot{\bm d}_{\alpha})/|{\bm z}_{\alpha}|^2\=({\bm z}_{\alpha}\cdot\bm{\mathcal{M}}\cdot{\bm z}_{\alpha})/|{\bm z}_{\alpha}|^2\=({\bm d}_{\alpha}\cdot\bm{\mathcal{M}}^{-1}\cdot{\bm d}_{\alpha})/({\bm d}_{\alpha}\cdot\bm{\mathcal{M}}^{-2}\cdot{\bm d}_{\alpha})$ is a mesoscopic elastic stiffness. It involves both a linear susceptibility related to $\bm{\mathcal{M}}^{-1}$ and a nonlinear one related to $\bm{\mathcal{M}}^{-2}\!\equiv\!\bm{\mathcal{M}}^{-1}\cdot\bm{\mathcal{M}}^{-1}$. The mean value $\langle\kappa_{\rm dipole}\rangle$ has been shown to be a characteristic stiffness (frequency squared) scale that is fundamentally related to soft glassy (nonphononic) modes~\cite{Lerner2018-me,Rainone2020-jo,Rainone2020-gs}.

%%%%%%%%%%%%%%%%%%%%%%%%%%%%%%%%%%%%%%%%%%%%%%%%%%%%%%%%%%%%
%%%%%%%%%%%%%%%%%%%%%%%%%%%%%%%%%%%%%%%%%%%%%%%%%%%%%%%%%%%%

Inspired by these developments, we aim at constructing an analogous micromechanical response quantity in the above-defined mean-field model and study its properties. This undertaking may appear at first sight non-trivial, if not impossible. The point is that unlike finite-dimensional glasses, the mean-field model is fully connected and lacks any spatial structure. Yet, as stressed above, the model has been shown to feature strongly localized low-frequency modes whose statistical properties mirror those of the corresponding objects in finite-dimensional glasses. In fact, as shown below, such analogous micromechanical objects can be rigorously defined through the linear response to force monopoles. Specifically, since the mean-field model lacks translational symmetry, the leading-order local perturbations that are analogous to dipolar perturbations in finite-dimensional glasses are force monopoles.

This physical and mathematical analogy, to be spelled out next, is strong and the resulting micromechanical response quantity is shown to encode rich and intriguing information about the model's underlying glassy potential-energy landscape. Consider the application of a local force monopole ${\bm f}_i$ to the $i^{\rm \tiny th}$ oscillator, i.e., ${\bm f}_i$ is an $N$-dimensional force vector with a unity $i^{\rm \tiny th}$ entry and zero elsewhere (for $j\!\ne\!i$). Denoting by ${\bm z}_i$ the resulting linear response field, the corresponding stored energy is $\frac{1}{2}\kappa_i|{\bm z}_i|^2$, where $\kappa_i$ is the $i^{\rm \tiny th}$ monopole stiffness. Since ${\bm z}_i\=\bm{\mathcal{M}}^{-1}\!\cdot\!{\bm f}_i$, where $\bm{\mathcal{M}}$ is the model's Hessian, one obtains
\begin{equation}
    \kappa_i = \frac{{\bm z}_i\cdot{\bm f}_i}{{\bm z}_i\cdot{\bm z}_i} = \frac{{\bm f}_i \cdot\bm{\mathcal{M}}^{-1}\cdot {\bm f}_i}{{\bm f}_i \cdot \bm{\mathcal{M}}^{-2}\cdot {\bm f}_i}=\frac{(\mathcal{M}^{-1})_{ii}}{(\mathcal{M}^{-2})_{ii}} > 0 \ ,
\label{eq:monopole_stiffness}
\end{equation}
where positivity is ensured since $\bm{\mathcal{M}}$ is positive definite.

The model's monopole stiffness $\kappa_i$ is analogous to the above-discussed dipolar stiffness $\kappa_{\rm dipole}^{(\alpha)}$, obtained in finite-dimensional glasses. Both quantities are defined through a concrete response protocol, where one applies a local perturbation and measures the resulting force-displacement relation, which makes it --- at least in principle --- accessible to experiments (e.g., via small-scale ``pinching'' protocols) and directly comparable across different models and computer simulations. An example of a monopole response in the mean-field model, embedded in 2D, is presented in Fig.~\ref{fig_monopole_short:Fig1}b. Despite the all-to-all nature of the interactions in the mean-field model, the observed monopole response reveals a signature of localization, in analogy to the 2D dipole response shown in Fig.~\ref{fig_monopole_short:Fig1}a, see caption for more details.

\begin{figure}[t]
    \centering
    \includegraphics[width=0.9\columnwidth]{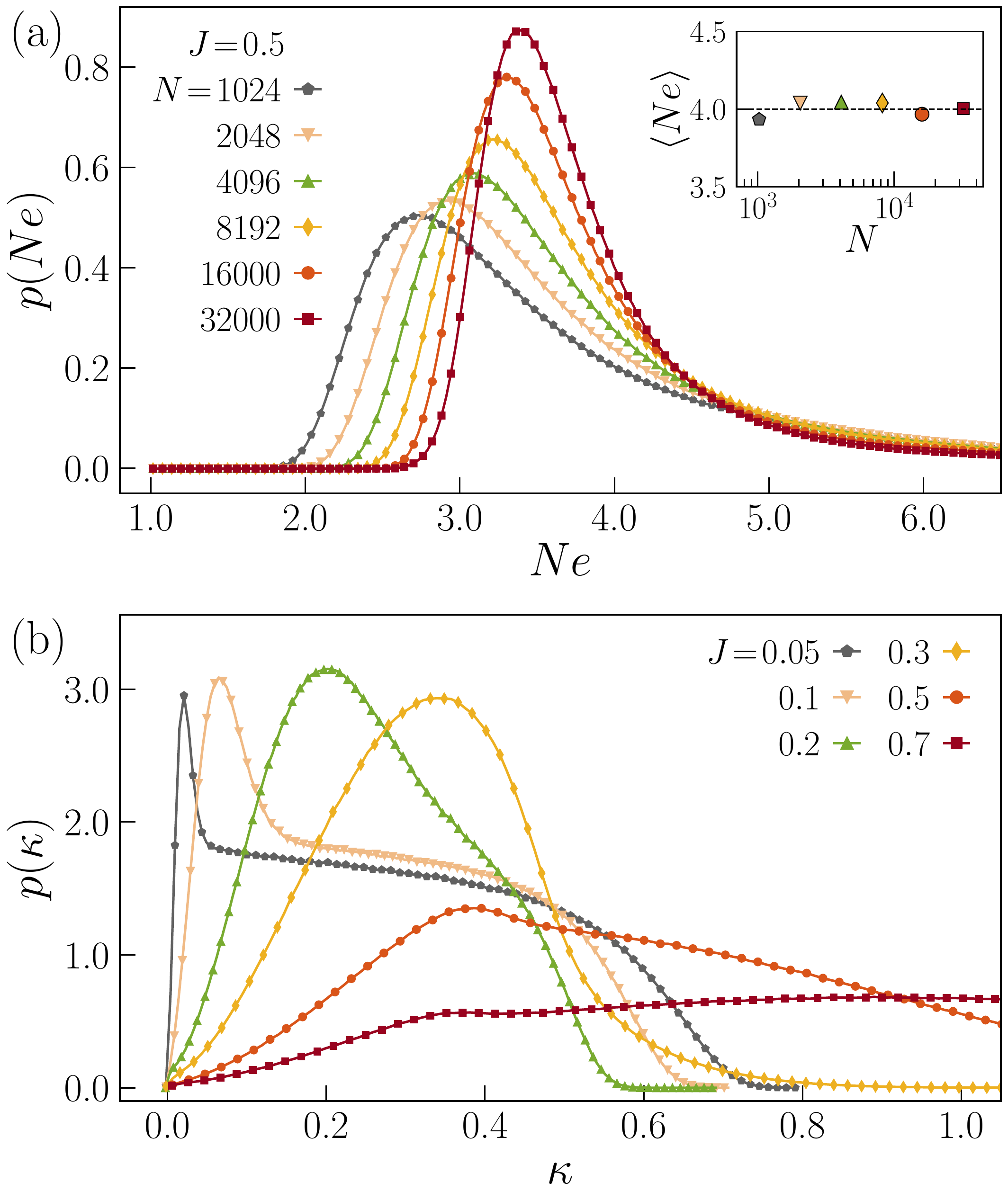}
    \caption{(a) The distribution $p(Ne)$ of the participation number $Ne$ of the monopole response $\bm z$ for $J\!=\!0.5$ and various system sizes $N$ (see legend). (inset) The mean participation number $\langle Ne\rangle$ vs.~$N$. The superimposed horizontal dashed line indicates an $N$-independent value, herein $\langle Ne\rangle\!=\!4$. (b) The distribution $p(\kappa)$ of monopole stiffnesses for $N\!=\!8192$ and various $J$ values (see legend). See text for definitions and discussion.}
    \label{fig_monopole_short:Fig2}
\end{figure}

To further explore the close correspondence between the mean-field monopole response and the dipole response in finite-dimensional glasses~\cite{Lerner2018-me,Rainone2020-jo,Rainone2020-gs}, we study $\kappa_i$ and ${\bm z}_i$ computed at minima of the Hamiltonian in Eq.~\eqref{eq:Hamiltonian} for various system sizes $N$ and disorder levels $J$. First, in Fig.~\ref{fig_monopole_short:Fig2}a, we plot the distribution $p(Ne)$, where $Ne$ is the participation number associated with the monopole response ${\bm z}$. $p(Ne)$ features an $N$-independent mean value $\langle Ne\rangle\!\simeq\!4$, as highlighted in the inset, indicating strong localization similarly to quasi-localized soft modes in computer glasses (i.e., $e\!\sim\!\mathcal{O}(1/N)$, despite the fully-connected nature of the model). Second, the monopole stiffness distribution $p(\kappa)$ presented in Fig.~\ref{fig_monopole_short:Fig2}b is pseudogapped, i.e., it vanishes smoothly as $\kappa\!\to\!0$, over a broad range of $J$ values. Since the initial distribution $p(k)$ is strictly positive at $k\!=\!0$, this pseudogap --- that indicates marginality --- is emergent. These two properties, i.e., the strong localization of linear responses to force monopoles and the marginality of their associated stiffness, closely parallel the behavior of dipole responses in finite-dimensional glasses~\cite{Lerner2018-me,Rainone2020-jo,Rainone2020-gs}, and thus provide direct preliminary support to the physical analogy between these micromechanical observables.
%%%%%%%%%%%%%%%%%%%%%%%%%%%%%%%%%%%%%%%%%%%%%%%%%%%%%%%%%%%%
%%%%%%%%%%%%%%%%%%%%%%%%%%%%%%%%%%%%%%%%%%%%%%%%%%%%%%%%%%%%

\section{A\lowercase{ mean-field identity}}
\label{sec:identity}

Our next goal is to use Eq.~\eqref{eq:monopole_stiffness}, along with standard Random Matrix Theory techniques~\cite{Bun2017-xj}, to derive a relation between the local monopole stiffness and global susceptibilities in the framework of the mean-field model of Eq.~\eqref{eq:Hamiltonian} (with no external force, i.e., $f(t)\=0$). The technical details of the derivation are provided in Appendix~\ref{appendix:mean-field_identity}, here we just delineate it and focus on the main result.

Since Eq.~\eqref{eq:monopole_stiffness} contains the diagonal elements of (different orders of) the inverse of the Hessian $\bm{\mathcal{M}}$, we first evaluate the elements of the latter using Eq.~\eqref{eq:Hamiltonian}, obtaining
\begin{equation}
    \mathcal{M}_{ij} = \delta_{ij}(k_i+\tfrac{1}{2}x_i^2) + J_{ij} \,,
\label{eq:Hessian}
\end{equation}
where no summation convention is used. $\mathcal{M}_{ij}$ is a sum of a diagonal matrix (first term) and a random matrix (second term), which is a member of the Gaussian Orthogonal Ensemble (GOE). Each diagonal element is in itself a sum of the initial stiffness $k_i$ and the squared displacement $\tfrac{1}{2}x_i^2$ induced by minimizing the Hamiltonian.

To proceed, we define the resolvent $\bm{G}(z)\!\equiv\!(z\bm{I}-\bm{\mathcal{M}})^{-1}$ associated with the Hessian $\bm{\mathcal{M}}$, where $z$ is in general a complex variable, which can be used to express the monopole stiffness as $\kappa_i\=(\mathcal{M}^{-1})_{ii}/(\mathcal{M}^{-2})_{ii}\=G_{ii}(0)/G'_{ii}(0)$,
% \begin{equation}
%     \kappa_i = \frac{(\mathcal{M}^{-1})_{ii}}{(\mathcal{M}^{-2})_{ii}} = \frac{G_{ii}(0)}{G'_{ii}(0)}\ ,
% \label{eq:resolvent_stiffness}
% \end{equation}
with $G'_{ii}(0)\!\equiv\!dG_{ii}(z\=0)/dz$. Assuming then that there is no statistical correlation between the Hessian diagonal and off-diagonal elements, and using some self-consistency Random Matrix Theory relations (see Appendix~\ref{appendix:mean-field_identity} for details), we obtain
\begin{equation}
    \kappa_i  = \frac{\left(k_i - J^2\chi\right) +\tfrac{1}{2}x_i^2}{1 + J^2\chi_{_{\rm SG}}} > 0 \ .
\label{eq:mean-field_identity}
\end{equation}
Here, $\chi\!\equiv\!N^{-1}{\rm Tr}(\bm{\mathcal{M}}^{-1})$ and $\chi_{_{\rm SG}}\!\equiv\!N^{-1}{\rm Tr}(\bm{\mathcal{M}}^{-2})$ are the zero-temperature linear and spin-glass (first nonlinear) global susceptibilities, respectively.

Equation~\eqref{eq:mean-field_identity} is a central result of this work. It relates the linear response to a force monopole applied to the $i^{\rm \tiny th}$ oscillator, whose corresponding stiffness is given in Eq.~\eqref{eq:monopole_stiffness}, and the diagonal element $k_i+\tfrac{1}{2}x_i^2$ (also associated with the $i^{\rm \tiny th}$ oscillator) to the global susceptibilities $\chi(J)$ and $\chi_{_{\rm SG}}(J)$, both multiplied by $J^2$. The fact that entirely local properties, such as $k_i$, $x_i$ and $\kappa_i$, are related to global susceptibilities is remarkable, yet it is a manifestation of the mean-field, fully-connected nature of the model.

The mean-field identity in Eq.~\eqref{eq:mean-field_identity} is expected to be asymptotically exact in the $N\!\to\!\infty$ limit and to be subjected to finite-$N$ fluctuations. The origin of latter is the existence of a finite minimal frequency $\omega_{\rm min}(N)$ in ${\cal D}_{\!_{\rm G}\!}(\omega)$, which vanishes in the $N\!\to\!\infty$ limit. A finite $\omega_{\rm min}(N)$ affects the convergence of global susceptibilities, which introduces finite-size effects. These are in fact anomalous, as discussed in more detail in Appendix~\ref{appendix:mean-field_identity}, where the validity of Eq.~\eqref{eq:mean-field_identity} and its convergence with increasing $N$ are demonstrated. In addition, some of the corollaries of Eq.~\eqref{eq:mean-field_identity} are extensively discussed in Appendix~\ref{appendix:related_identities}.

An immediate and important implication of Eq.~\eqref{eq:mean-field_identity} is that in order for the micromechanical response quantity $\kappa_i$ to be well-defined and finite, the spin-glass susceptibility $\chi_{_{\rm SG}}(J)\=\int\!\omega^{-4}\,{\cal D}_{\!_{\rm G}\!}(\omega)\,d\omega$ has to be finite. In turn, the latter condition implies that ${\cal D}_{\!_{\rm G}\!}(\omega)$ should be pseudogapped, with a power-law tail exponent larger than $3$. Indeed, as stressed above, the model universally features ${\cal D}_{\!_{\rm G}\!}(\omega)\!\sim\!\omega^4$, which self-consistently satisfies the condition.

To highlight the previous point, we note that models that feature a pseudogapped VDoS with a power-law tail exponent smaller than $3$, e.g., many known models featuring ${\cal D}_{\!_{\rm G}\!}(\omega)\!\sim\!\omega^2$ in the $\omega\!\to\!0$ limit, are expected not to possess a well-defined monopole response due to their divergent spin-glass susceptibility. This is explicitly demonstrated in Appendix~\ref{appendix:3spin_model} for the spherical 3-spin model, which features ${\cal D}_{\!_{\rm G}\!}(\omega)\!\sim\!\omega^2$ (with delocalized modes), leading to a divergent $\chi_{_{\rm SG}}$ and ill-defined $\kappa_i$'s.

\section{C\lowercase{omputing vibrational spectra from monopole responses}}
\label{sec:VDoS}

We concluded the previous section by highlighting a self-consistency relation between the monopole stiffness $\kappa_i$ and the low-$\omega$ tail of ${\cal D}_{\!_{\rm G}\!}(\omega)$. Here, we explore stronger relations by asking whether the statistics of $\{\kappa_i\}$ can be used to compute ${\cal D}_{\!_{\rm G}\!}(\omega)$ over the entire frequency range.

A central quantity in this context is the normalized trace of the resolvent $\mathfrak{g}_{_{_{\bm{\mathcal{M}}}}}\!(z)\!\equiv\!N^{-1}{\rm Tr}[\bm{G}(z)]\=N^{-1}{\rm Tr}[(z\bm I \!-\!\bm{\mathcal{M}})^{-1}]$. Using standard Random Matrix Theory, it can be shown that $\mathfrak{g}_{_{_{\bm{\mathcal{M}}}}}\!(z)$ satisfies the following self-consistency relation (see Appendix~\ref{appendix:VDoS_reconstruction} for details)
\begin{equation}
    \mathfrak{g}_{_{_{\bm{\mathcal{M}}}}}\!(z)=N^{-1}\sum_i\frac{1}{z-a_i-J^2\,\mathfrak{g}_{_{_{\bm{\mathcal{M}}}}}\!(z)}\,,
\label{eq:resolvent_self-consistency}
\end{equation}
where $a_i\!\equiv\!k_i\!+\!\tfrac{1}{2}x_i^2$ is a diagonal element of the Hessian, see Eq.~\eqref{eq:Hessian}. This relation suggests that the statistics of $\{a_i\}$ can be used to solve Eq.~\eqref{eq:resolvent_self-consistency} to obtain the continuous complex-variable function $\mathfrak{g}_{_{_{\bm{\mathcal{M}}}}}\!(z)$. Once $\mathfrak{g}_{_{_{\bm{\mathcal{M}}}}}\!(z)$ is at hand, ${\cal D}_{\!_{\rm G}\!}(\omega)$ can be readily obtained as the eigenvalue $\lambda\=\omega^2$ density satisfies $\rho_{_{_{\bm{\mathcal{M}}}}}\!(\lambda)\={\pi}^{-1}\!\lim_{\eta\to 0^{+}}\!\Im\left[\mathfrak{g}_{_{_{\!\bm{\mathcal{M}}}}}\!(\lambda-i\eta)\right]$, from which ${\cal D}_{\!_{\rm G}\!}(\omega)\=2\omega\rho_{_{_{\!\bm{\mathcal{M}}}}}\!(\omega^2)$ follows. The suggested procedure amounts to the following sequence of transformations: $p(\kappa)\!\to\!p(a)\!\to\!\mathfrak{g}_{_{_{\!\bm{\mathcal{M}}}}}\!(z)\!\to\!{\cal D}_{\!_{\rm G}\!}(\omega)$.

%%%%%%%%%%%%%%%%%%%%%%%%%%%%%%%%%%%%%%%%%%%%%%%%%%%%%%%%%%%%
\begin{figure}[t]
    \centering
    \includegraphics[width=1\columnwidth]{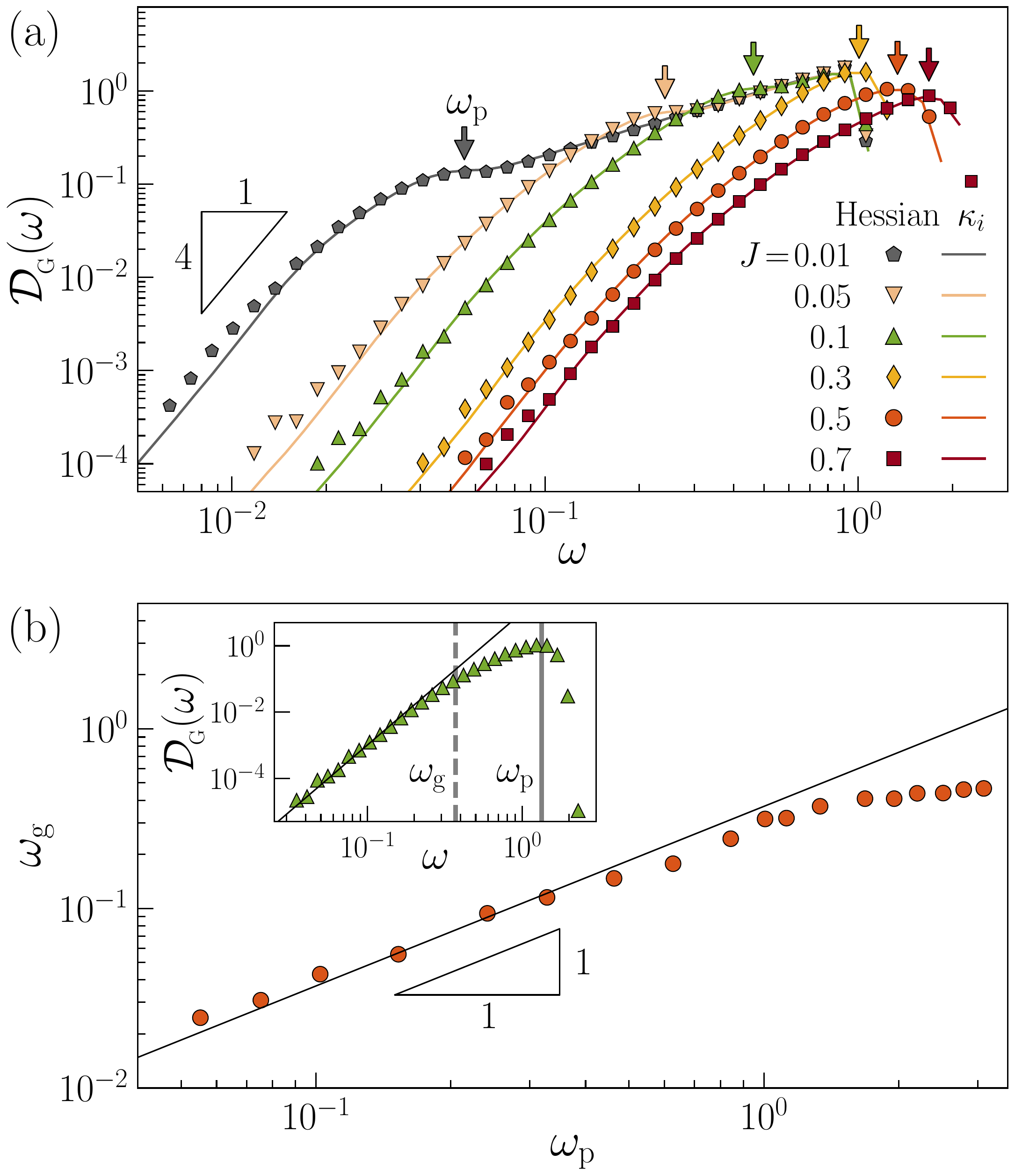}
    \vspace{-0.4cm}
    \caption{(a) The nonphononic VDoS ${\cal D}_{\!_{\rm G}\!}(\omega)$, obtained both from diagonalizing the Hessian (marked in legend as ``Hessian''; filled symbols) and based on the reconstruction from monopole stiffness statistics (marked in legend as ``$\kappa_i$''; solid lines), see text, for various $J$ (see legend). The local maximum of ${\cal D}_{\!_{\rm G}\!}(\omega)$, termed the intrinsic boson-peak frequency, is marked on each curve (see downward-pointing arrows) and denoted as $\omega_{\rm p}$. (b) $\omega_{\rm p}(J)$ vs.~the upper-cutoff frequency $\omega_{\rm g}(J)$ of the $\sim\!\omega^4$ low-frequency tail, plotted for $0.01\!\leq\!J\!\leq\!2$ (from bottom-left to top-right). The superimposed solid line indicates $\omega_{\rm p}\!\sim\!\omega_{\rm g}$. (inset) Demonstrating the extraction of $\omega_{\rm g}$, here for $J\!=\!0.5$. The solid line is a fit to ${\cal D}_{\!_{\rm G}\!}(\omega)\!=\!A_{\rm g}\omega^4$ and the vertical dashed line is the extracted $\omega_{\rm g}$ (see Appendix~\ref{appendix:omega_g} for the adopted operational definition). The vertical solid line corresponds to $\omega_{\rm p}$ of panel (a). We use $M\!=\!2000$ realizations of the quenched disorder at $N\!=\!8192$ for each $J$ value. See text for discussion. In Appendix~\ref{appendix:VDoS_reconstruction} and~\ref{appendix:omega_g}, we detail the computation of the solid lines in (a), and the extraction of $\omega_{\rm g}$ in (b).}
    \label{fig_monopole_short:Fig3}
\end{figure}
%%%%%%%%%%%%%%%%%%%%%%%%%%%%%%%%%%%%%%%%%%%%%%%%%%%%%%%%%%%%

To implement this procedure, one uses $\{\kappa_i\}$ to obtain $\langle \kappa^{-1}\rangle\!\equiv\!N^{-1}\sum_i\kappa_i^{-1}$ and likewise $\langle \kappa^{-2}\rangle$. Equation~\eqref{eq:mean-field_identity} is then used to obtain the susceptibilities $\chi$ and $\chi_{_{\rm SG}}$, and finally $a_i$ is expressed in terms of $\kappa_i$, $\langle \kappa^{-1}\rangle$ and $\langle \kappa^{-2}\rangle$ (see Appendix~\ref{appendix:related_identities}--\ref{appendix:VDoS_reconstruction} for details). Solving then Eq.~\eqref{eq:resolvent_self-consistency} numerically as described in Appendix~\ref{appendix:VDoS_reconstruction}, the resulting ${\cal D}_{\!_{\rm G}\!}(\omega)$ is plotted in Fig.~\ref{fig_monopole_short:Fig3}a (continuous lines) for a broad range of $J$ values (see legend). The validity of the procedure is demonstrated by superposing (discrete symbols) the VDoS independently obtained by a direct diagonalization of the Hessian over the very same ensemble of disorder realizations per $J$, revealing excellent quantitative agreement.

The intrinsic boson peak frequency, denoted as $\omega_{\rm p}$ (see Fig.~\ref{fig_monopole_short:Fig3}a), corresponds to the local maximum of ${\cal D}_{\!_{\rm G}\!}(\omega)$, as discussed in~\cite{Moriel2024-wp,Suda2025-kl}. It is ``intrinsic" because it characterizes the nonphononic VDoS ${\cal D}_{\!_{\rm G}\!}(\omega)$, while the conventional boson peak only indirectly probes the properties of the nonphononic VDoS, as it involves a ratio between the total and Debye's VDoS of phonons~\cite{Ramos2022-nf}. While the two are related, there are reasons to believe that the former is a more basic physical quantity than the latter~\cite{Parisi2003-ex,Gurevich2003-oo,Parshin2007-od}. Note that the local maximum of ${\cal D}_{\!_{\rm G}\!}(\omega)$ at $\omega_{\rm p}$ does not identify with the global maximum below some value of $J$ (see Fig.~\ref{fig_monopole_short:Fig3}a). In this case, $\omega\!>\!\omega_{\rm p}$ features a predominantly linear VDoS, ${\cal D}_{\!_{\rm G}\!}(\omega)\!\sim\!\omega$, which corresponds to a part of initial liquid-like $p(k)\=k_0^{-1}$ distribution that did not undergo interaction-induced ``reconstruction'' during the minimization of the Hamiltonian, as discussed in~\cite{Rainone2021-am,Moriel2024-wp}.

$\omega_{\rm p}$ is an important frequency scale in the model, and likewise in glasses~\cite{Ramos2022-nf}. It is natural to ask whether it is the only characteristic frequency of ${\cal D}_{\!_{\rm G}\!}(\omega)$. It is clear that the quartic tail also involves a frequency scale~\cite{Rainone2020-jo}. To see this, note that the non-universal, disorder-dependent prefactor $A_{\rm g}(J)$ in ${\cal D}_{\!_{\rm G}\!}(\omega)\=A_{\rm g}(J)\,\omega^4$ is of dimension of frequency to the power of $-5$ and hence incorporates a frequency scale, to be denoted as $\omega_{\rm g}$, which
% . That is, the VDoS tail actually takes the form ${\cal D}_{\!_{\rm G}\!}(\omega)\!\sim\!\omega^4\!/\omega_{\rm g}^5$, where $\omega_{\rm g}$
sets the upper cutoff on the power-law scaling (similarly to Debye's frequency for phonons~\cite{Kittel2004}). The question then is whether $\omega_{\rm g}(J)$ scales with $\omega_{\rm p}(J)$ as $J$ is varied. This is tested in Fig.~\ref{fig_monopole_short:Fig3}b, demonstrating that $\omega_{\rm g}(J)$ systematically and monotonically varies with $\omega_{\rm p}(J)$, rather linearly over a broad range, which implies that the nonphononic VDoS is predominantly controlled by a single frequency scale. A similar relation has been proposed for glasses~\cite{Lerner2018-me,Rainone2020-jo,Rainone2020-gs}, though it has not been quantitatively and systematically tested so far.

\section{A\lowercase{ stabilization criterion and a characteristic stiffness/frequency scale}}
\label{sec:freq_scale}

The existence of a dominant characteristic frequency scale in ${\cal D}_{\!_{\rm G}\!}(\omega)$, demonstrated above, raises the question of whether it is possible, and if so how, to extract this basic quantity directly from the monopole stiffnesses $\{\kappa_i\}$. It is natural to expect that this is possible as it was demonstrated in Fig.~\ref{fig_monopole_short:Fig3}a that $\{\kappa_i\}$ can be used to compute ${\cal D}_{\!_{\rm G}\!}(\omega)$. In addition, it has been suggested in the context of finite-dimensional computer glasses that the mean dipole stiffness $\langle\kappa_{\rm dipole}\rangle$ is a proxy of $\omega_{\rm g}$, i.e., $\sqrt{\langle\kappa_{\rm dipole}\rangle}\!\sim\!\omega_{\rm g}$ when the state of disorder of a glass is varied. As the monopole stiffness is the analogous quantity in the mean-field model, we first consider $\sqrt{\langle\kappa\rangle}$ as an estimate for the characteristic frequency. We found that $\omega_{\rm p}$ does not scale with $\sqrt{\langle\kappa\rangle}$. Likewise, the reduced distribution $p(\kappa/\langle\kappa\rangle)$ does not collapse on a master curve when $J$ is varied, as is evident from Fig.~\ref{fig_monopole_short:Fig2}b.

These results may indicate that either a different local stiffness should be defined or that only a subpopulation of the oscillators, and their associated monopole stiffnesses $\{\kappa_i\}$, control the physically relevant characteristic frequency scale. A naive alternative definition of a local stiffness $\tilde{\kappa}_i$ would be through the response to displacing the $i^{\rm \tiny th}$ oscillator by amount $\delta{x}_i$, while keeping all other oscillators fixed. That is, we define $\delta{\bm x}_i$ as an $N$-dimensional displacement vector with an $i^{\rm \tiny th}$ entry of magnitude $\delta{x}_i$ and zero elsewhere (for $j\!\ne\!i$). The stiffness associated with the response to $\delta{\bm x}_i$ then reads $\tilde{\kappa}_i\!\equiv\!(\delta{\bm x}_i\!\cdot\!\bm{\mathcal{M}}\!\cdot\!\delta{\bm x}_i)/|\delta{\bm x}_i|^2\={\cal M}_{ii}$. That is, $\tilde{\kappa}_i\=k_i+\tfrac{1}{2}x_i^2$ is nothing but the diagonal element of the Hessian, see Eq.~\eqref{eq:Hessian}, which also appears in the numerator on the right-hand-side (RHS) of the mean-field identity in Eq.~\eqref{eq:mean-field_identity}. We found that $\omega_{\rm p}$ also does not scale with $\sqrt{\langle\tilde\kappa\rangle}$.

We therefore explore the possibility that only a subset of $\{\kappa_i\}$ controls the physically relevant characteristic frequency scale. To this aim, let us reexamine the mean-field identity in Eq.~\eqref{eq:mean-field_identity} and ask how the positivity of $\kappa_i$ comes about. First, recall that $\chi_{_{\rm SG}},\chi\!>\!0$, implying that the denominator of Eq.~\eqref{eq:mean-field_identity} is positive and its numerator features a single negative contribution, $-J^2\chi\!<\!0$. Then, since $k_i$ is uniformly distributed over the interval $[0,1]$ (recall that we set $k_0\=1$), there exists $\kappa_i$ values for which $k_i\!-\!J^2\chi\!<\!0$. For this subpopulation of oscillators, $\kappa_i\!>\!0$ is achieved only due to sufficiently large quench-generated landscape displacement $\frac{1}{2}x_i^2$. This is to be contrasted with oscillators for which $k_i\!-\!J^2\chi\!>\!0$, whose stability/positivity does not depend on the magnitude of $\frac{1}{2}x_i^2$. We therefore propose that the sign of $k_i\!-\!J^2\chi$ separates the entire set of oscillators into two subpopulation, those that are ``stabilized'' (by the displacements $\frac{1}{2}x_i^2$), i.e., characterized by $k_i-J^2\chi\!<\!0$, and the complementary subpopulation characterized by $k_i-J^2\chi\!>\!0$.

%%%%%%%%%%%%%%%%%%%%%%%%%%%%%%%%%%%%%%%%%%%%%%%%%%%%%%%%%%%%
\begin{figure}[t]
    \centering
    \includegraphics[width=0.95\columnwidth]{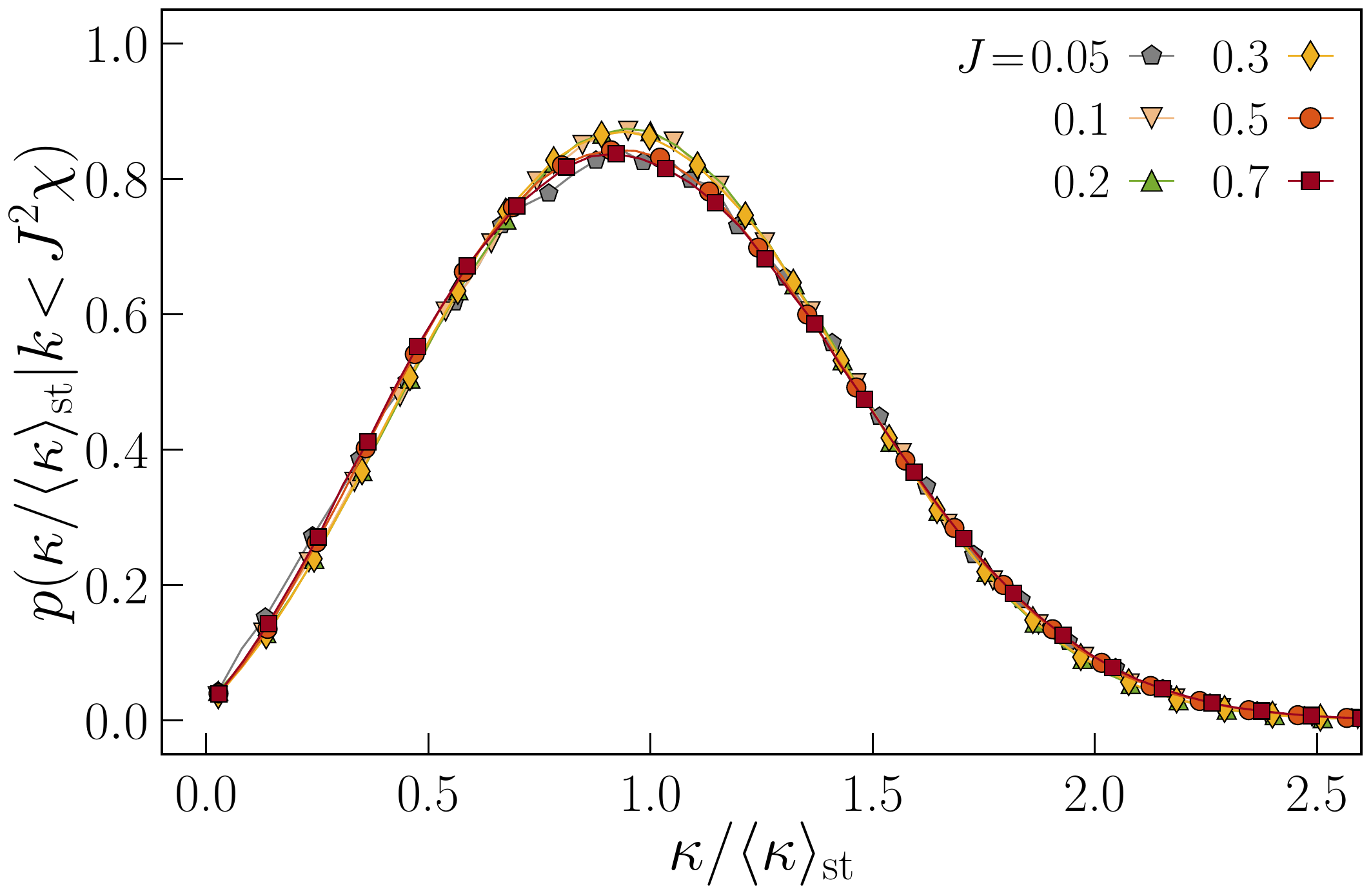}
    \vspace{-0.4cm}
    \caption{The collapsed distributions $p(\kappa/\langle\kappa\rangle_{\rm st}|k\!<\!J^2\chi)$ of stabilized-sector monopole stiffness for various $J$ values and $N\!=\!8192$ (see legend), where $\langle \kappa\rangle_{\rm st}(J)$ is the mean stabilized-sector stiffness. See text for definitions and discussion.}
    \label{fig_monopole_short:Fig4}
\end{figure}
%%%%%%%%%%%%%%%%%%%%%%%%%%%%%%%%%%%%%%%%%%%%%%%%%%%%%%%%%%%%

While the above subset dichotomy emerges from micromechanical stability arguments for local monopole responses, they also echo related distinctions recently appeared in~\cite{Folena2022-tn} and in the so-called ``reconstruction picture" for the glassy nonphononic spectra~\cite{Gurevich2003-oo,Parshin2007-od}. The same distinction also emerges in the scaling theory of~\cite{Rainone2021-am}. The crux of these approaches is that degrees of freedom softer than the interaction-induced stiffness shift $J^2\chi$ reconstruct/self-organize to form glassy nonphononic modes. Taken together, we expect key properties of nonphononic modes to be encoded in the monopole responses of this subset of oscillators.

To test this idea, we consider the monopole stiffness distribution of the $k_i\!<\!J^2\chi$ oscillators, to be termed hereafter the ``stabilized sector", i.e., the conditional distribution $p(\kappa|k\!<\!J^2\chi)$. We show in Fig.~\ref{fig_monopole_short:Fig4} that this conditional distribution collapses across a range of quenched disorder level $J$, when rescaled by the mean stabilized-sector stiffness
\begin{equation}
    \langle \kappa \rangle_{\rm st}(J) \equiv \int \kappa\, p(\kappa|k<J^2\chi)\,d\kappa\ ,
\label{eq:stabilized-sector_mean}
\end{equation}
where the subscript ``st" stands for ``stabilized". This shape-invariance of $p(\kappa|k\!<\!J^2\chi)$ across $J$ should be contrasted with the lack of a corresponding shape-invariance of the unconditional distribution $p(\kappa)$, plotted in Fig.~\ref{fig_monopole_short:Fig2}b. It shows that $\langle\kappa\rangle_{\rm st}$ is the characteristic stiffness scale controlling the stabilized-sector statistics.

%%%%%%%%%%%%%%%%%%%%%%%%%%%%%%%%%%%%%%%%%%%%%%%%%%%%%%%%%%%%
\begin{figure}[t]
    \centering
    \includegraphics[width=0.95\columnwidth]{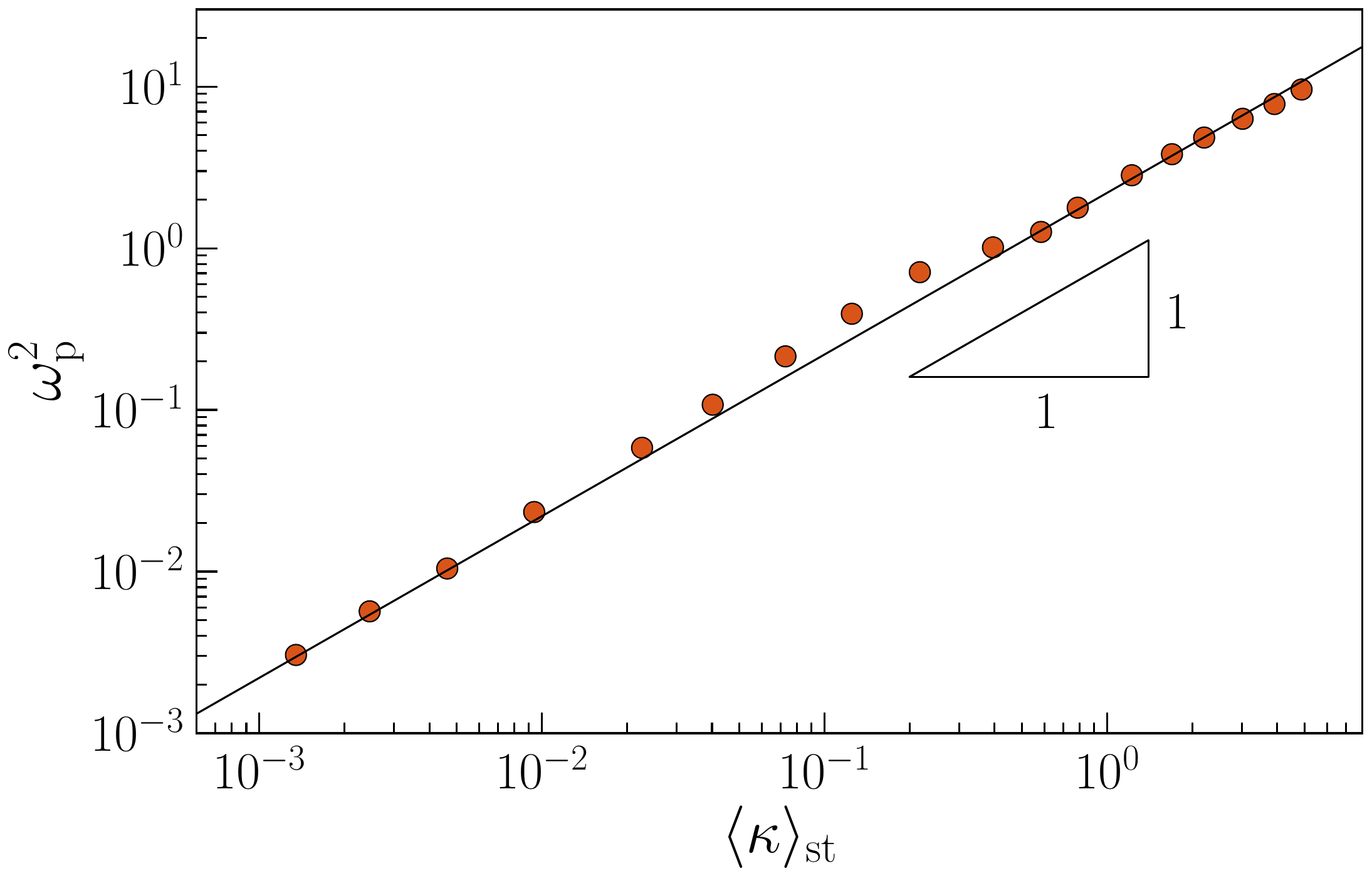}
    \caption{The characteristic stiffness scale $\omega^2_{\rm p}(J)$ of the nonphononic modes vs.~$\langle\kappa\rangle_{\rm st}(J)$ for $N\!=\!8192$ and $0.01\!\leq\!J\!\leq\!2$ (from bottom-left to top-right). The superimposed solid line corresponds to $\langle\kappa\rangle_{\rm st}(J)\!\sim\!\omega^2_{\rm p}(J)$.}
    \label{fig_monopole_short:Fig5}
\end{figure}
%%%%%%%%%%%%%%%%%%%%%%%%%%%%%%%%%%%%%%%%%%%%%%%%%%%%%%%%%%%%

With a characteristic stiffness scale $\langle\kappa\rangle_{\rm st}(J)$ at hand, we explore its relation to the intrinsic boson-peak frequency $\omega_{\rm p}(J)$ in the VDoS ${\cal D}_{\!_{\rm G}\!}(\omega)$, discussed in the context of Fig.~\ref{fig_monopole_short:Fig3}. In Fig.~\ref{fig_monopole_short:Fig5}, we plot one against the square of the other over the range $0.01\!\leq\!J\!\leq\!2$. The results establish the scaling relation
\begin{equation}
\langle\kappa\rangle_{\rm st} \sim \omega_{\rm p}^2 \ ,
\label{eq:kappa_w_p_scaling}
\end{equation}
demonstrating that $\sqrt{\langle\kappa\rangle_{\rm st}}$ indeed probes the frequency scale of the nonphononic modes. It also suggests a relation between a measure of mesoscopic elasticity and the boson-peak frequency, both of which are in principle measurable in laboratory glasses.

\section{C\lowercase{omparing mesoscopic and global elastic stiffnesses}}
\label{sec:global_vs_mesoscopic}

As noted above, the characteristic stiffness scale $\langle\kappa\rangle_{\rm st}$, in analogy to the characteristic dipolar stiffness scale in finite-dimensional glasses, may be regarded as probing ``mesoscopic elasticity'' in the mean-field model, as it corresponds to the collective response to a local force monopole. In computer glasses, the variation of the mesoscopic stiffness with the glass history (i.e., state of disorder) has been compared to the corresponding variation of its macroscopic counterpart, i.e., the continuum-level shear modulus~\cite{Lerner2018-me,Rainone2020-jo,Rainone2020-gs}. A generic outcome of such analyses is that the mesoscopic stiffness varies over a broader range compared to the macroscopic one~\cite{Lerner2018-me,Rainone2020-jo,Rainone2020-gs}, with potentially important implications for thermal glassy relaxation, in the context of the so-called ``elastic models'' of the glass transition~\cite{Dyre2006-ok,Ciamarra2026-mg}. Our goal here is to explore the corresponding comparison in the framework of the mean-field model.

To that aim, we need to define a global (``macroscopic'') elastic response coefficient in the mean-field model. Consider then a finite-dimensional glass subjected to a small, uniform shear stress of magnitude $\sigma$. The resulting linear elastic energy density (energy $U$ per volume $V$) takes the form $U/V\=\tfrac{1}{2}\sigma^2/G$, where the compliance $G^{-1}\=d^2(U/V)/d\sigma^2$ is the inverse of the shear modulus $G$. The direct analog of $\sigma$ in the mean-field model is $f$ in Eq.~\eqref{eq:Hamiltonian}, applied after the Hamiltonian $H$ is minimized. Note that $f$ is applied to all oscillators therein, whereas the local force monopole ${\bm f}_i$, extensively discussed above, is applied only to the $i^{\rm \tiny th}$ oscillator. Consequently, in analogy to the finite-dimensional case, we define $G^{-1}\!\equiv\!d^2(H/N)/df^2$ as the global compliance in the mean-field model. Using this definition and mechanical equilibrium, one obtains (see Appendix~\ref{appendix:global_modulus} for details)
% \begin{equation}
%     G = N\Big(\sum_{i,j}\mathcal{M}_{ij}^{-1}\Big)^{-1}
% \label{eq:modulus}
% \end{equation}
\begin{equation}
    G = N\Big(\bm 1\cdot\bm{\mathcal{M}}^{-1}\cdot \bm 1\Big)^{-1}\,,
\label{eq:modulus}
\end{equation}
where $\bm 1\!\equiv\!(1,...,1)$, for the global modulus $G$ in the mean-field model.

%%%%%%%%%%%%%%%%%%%%%%%%%%%%%%%%%%%%%%%%%%%%%%%%%%%%%%%%%%%%
\begin{figure}[t]
    \centering
    \includegraphics[width=0.95\columnwidth]{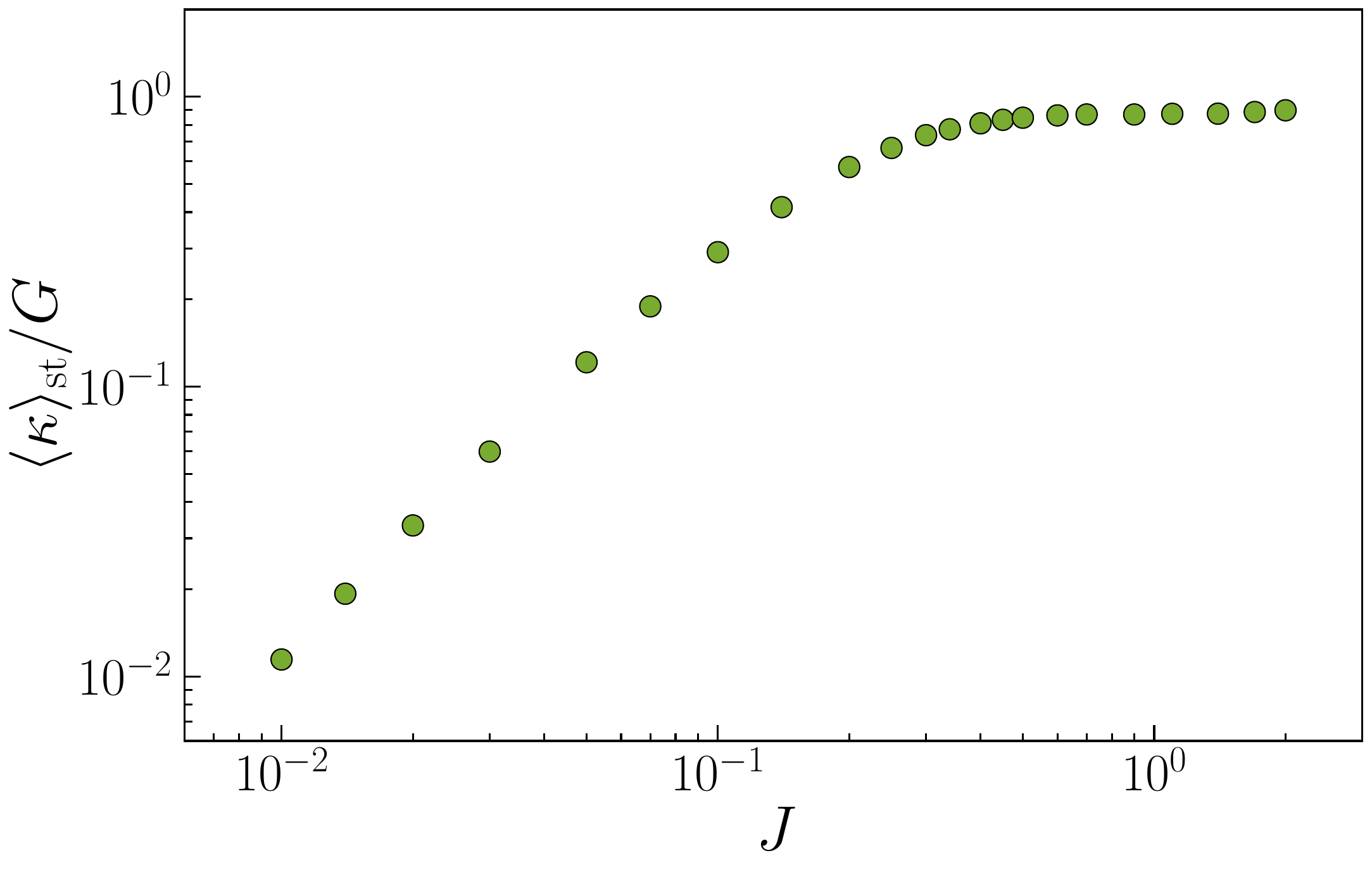}
    \caption{The ratio of the mesoscopic stiffness $\langle \kappa\rangle_{\rm st}$ and the global one $G$, for $0.01\!\le\!J\!\le\!2$ and $N\!=\!8192$. See text for definitions and discussion.}
    \label{fig_monopole_short:Fig6}
\end{figure}
%%%%%%%%%%%%%%%%%%%%%%%%%%%%%%%%%%%%%%%%%%%%%%%%%%%%%%%%%%%%

We then computed the global modulus $G(J)$ of Eq.~\eqref{eq:modulus} over a broad range of quenched disorder levels,
$0.01\!\leq\!J\!\leq\!2$. As we are interested in the range of variation of the latter compared to that of the mesoscopic stiffness $\langle \kappa\rangle_{\rm st}(J)$, already computed Fig.~\ref{fig_monopole_short:Fig5}, we plot the ratio $\langle\kappa\rangle_{\rm st}/G$ against $J$ in Fig.~\ref{fig_monopole_short:Fig6}. While each of these two quantities softens with decreasing $J$ (not shown individually), their ratio is constant at large $J$ and decreases with decreasing $J$ at smaller $J$ values, indicating that $\langle \kappa\rangle_{\rm st}(J)$ features a larger range of variation compared to $G(J)$, similarly to the corresponding observations in computer glasses~\cite{Lerner2018-me,Rainone2020-jo,Rainone2020-gs}.

\section{A\lowercase{nalogy to the effective potential}}
\label{sec:relation_to_effective_potential}

The intrinsic relation between the stabilized-sector of the force monopole response and the  stiffness/frequency scale of nonphononic modes, established in Sect.~\ref{sec:freq_scale}, indicates that the distinction based on the sign of $k_i\!-\!J^2\chi$ may have deeper implications both for the glassiness of the model in a broader context and for its relations to finite-dimensional glasses. Here, we explore possible relations between the monopole response and the effective potential in Eq.~\eqref{eq:eff_potential}. The stiffness of the uncoupled oscillator subjected to the random potential in Eq.~\eqref{eq:eff_potential} takes the form
\begin{equation}
    v''_{\rm eff}(x)=\left(k-J^2\chi^{(\rm RS)}\right)+\tfrac{1}{2}x^2  \ .
    \label{eq:stiffness_eff_potential}
\end{equation}
Moreover, the sign of $k_{\rm m}\!-\!J^2\chi^{(\rm RS)}$ (where $k_{\rm m}$ is the minimal stiffness of the initial $p(k)$ distribution, as defined above) played a crucial role in the equilibrium, zero-temperature RSB transition associated with Eq.~\eqref{eq:eff_potential}~\cite{Bouchbinder2021-dh,Folena2022-tn}.

\begin{figure}[t]
    \centering
    \includegraphics[width=0.95\columnwidth]{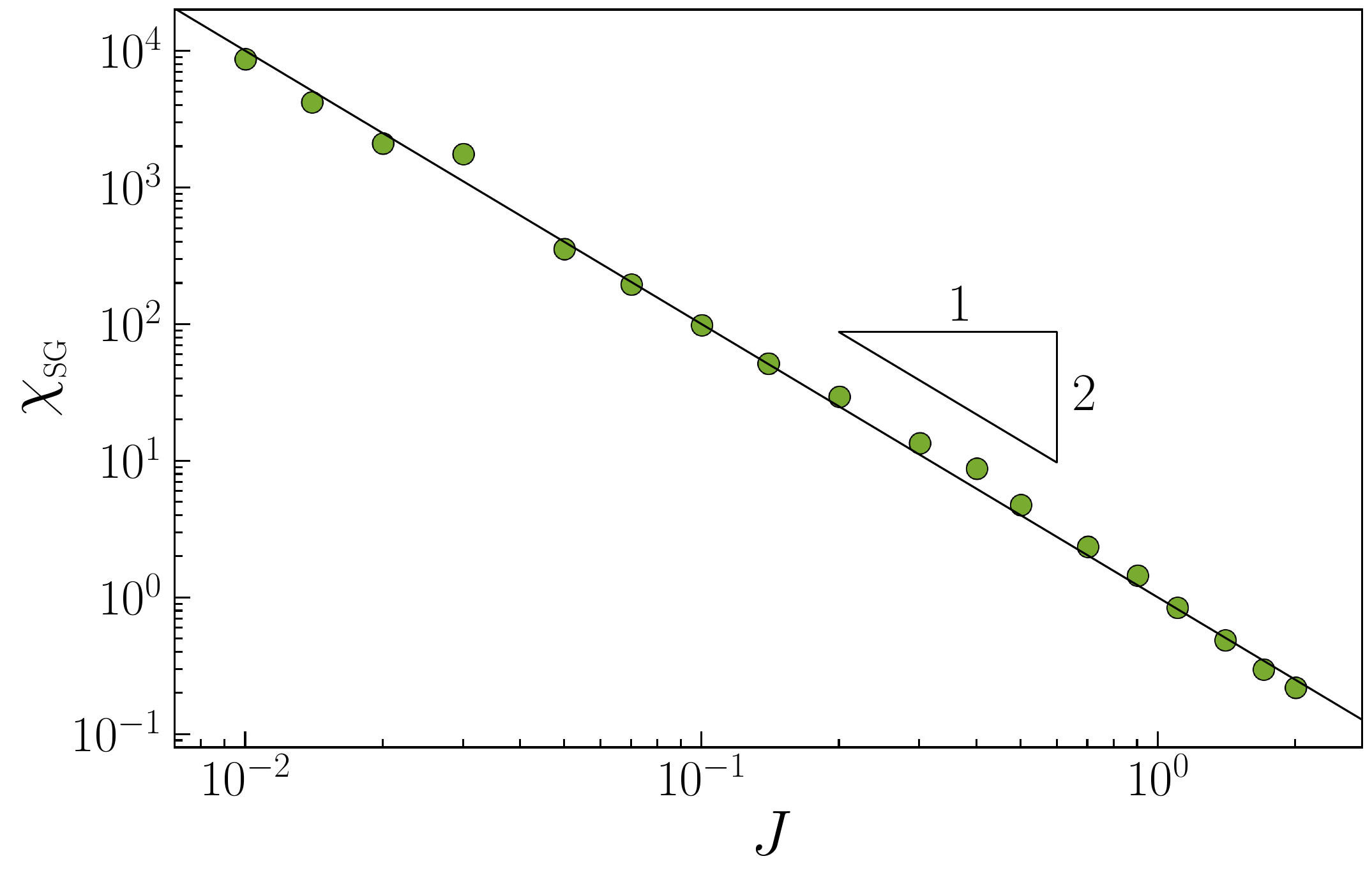}
    \caption{The spin-glass susceptibility $\chi_{_{\rm SG}}\!=\!N^{-1}{\rm Tr}(\bm{\mathcal{M}}^{-2})$ vs.~$J$ for $N\!=\!8192$. The superimposed solid line corresponds to $\chi_{_{\rm SG}}\!=\!J^{-2}$. See text for discussion.}
\label{fig_monopole_short:Fig7}
\end{figure}

This sign consideration of the shifted initial oscillator stiffness $k$ and more so the mathematical similarity between the effective stiffness in Eq.~\eqref{eq:stiffness_eff_potential} and the mean-field identity in Eq.~\eqref{eq:mean-field_identity} appear to suggest a deep correspondence between the force monopole response and the effective potential of Eq.~\eqref{eq:eff_potential}. This correspondence would appear stronger if $J^2\chi_{_{\rm SG}}(J)$ in the denominator of Eq.~\eqref{eq:mean-field_identity} is independent of $J$. Indeed, the relation $\chi_{_{\rm SG}}(J)\!\sim\!J^{-2}$ is demonstrated in Fig.~\ref{fig_monopole_short:Fig7}.
%Note that $J^2\chi_{_{\rm SG}}\!=\!1$ may be related to an analysis presented in~\cite{??} .
%%%%%%%%%%%%%%%%%%%%%%%%%%%%%%%%%%%%%%%%%%%%%%%%%%%%%%%%%%%%
%%%%%%%%%%%%%%%%%%%%%%%%%%%%%%%%%%%%%%%%%%%%%%%%%%%%%%%%%%%%

Consequently, Eqs.~\eqref{eq:mean-field_identity} and~\eqref{eq:stiffness_eff_potential} are essentially identical if the stiffness shift --- $J^2\chi$ in the former and $J^2\chi^{(\rm RS)}$ in the latter --- is basically the same in both. This correspondence may have important implications. First, it suggests that the mathematical structure of the effective potential Eq.~\eqref{eq:eff_potential}, derived in the RS phase approaching the equilibrium RSB transition in a certain parameter range of the model, remains valid deep inside the non-equilibrium RSB phase, in a parameter range where an equilibrium RS phase does not exist at all. These findings reinforce the theoretical suggestions of~\cite{Folena2022-tn}. Second, the correspondence suggests that the force monopole responses within the stabilized-sector constitute a concrete micromechanical realization of the relevant collective degrees of freedom of the model. Likewise, it lends additional support to the corresponding dipole response findings in finite-dimensional glasses~\cite{Lerner2018-me,Rainone2020-jo,Rainone2020-gs}.

\section{M\lowercase{ean-field force fluctuations and their relation to the boson peak}}
\label{sec:MF_force_fluctuations}

The correspondence to the effective potential focused so far on the curvature of $v_{\rm eff}(x)$, i.e., on the stiffness and its relation to the monopole response. Yet, the correspondence can be pushed further to include the random force fluctuations $\delta{\!f}$ in Eq.~\eqref{eq:eff_potential} (which do not contribute to $v''_{\rm eff}(x)$). That is, if we accept the correspondence between Eq.~\eqref{eq:mean-field_identity} and Eq.~\eqref{eq:stiffness_eff_potential} (and recall that $J^2\chi_{_{\rm SG}}\=1$, cf.~Fig.~\ref{fig_monopole_short:Fig7}), we can invoke force balance at minima of the Hamiltonian, which applies to any oscillator, to associate with it a force $\delta{\!f}$. Specifically, a force $\delta{\!f}_{\!i}$ is assigned to the $i^{\rm \tiny th}$ oscillator, characterized by stiffness $k_i$ and displacement $x_i$, according to $\delta{\!f}_{\!i}\=(k_i-J^2\chi)x_i+x_i^3/6$ (which is analogous to $v'_{\rm eff}(x^*)\=0$ in the language of the effective potential).

This mapping/assignment allows us to compute the distribution of the random force fluctuations $p(\delta{\!f})$. As mentioned above, the latter is expected to be Gaussian with zero mean in the RS phase. We find that indeed $\langle\delta{\!f}\rangle\=0$, which follows from symmetry, though $p(\delta{\!f})$ is not generally Gaussian, where deviations from Gaussianity appear near $\delta{\!f}\=0$ (see Appendix~\ref{appendix:random_force_fluctuation} for details). This observation seems to be related to the analysis of~\cite{Folena2022-tn} right beyond the RSB transition, where it was claimed that the probability density of $h_i^{(\rm loc)}\!\equiv\!\delta{\!f}_{\!i}+J^2\chi\,x_i$ in the stabilized sector, $k_i\!<\!J^2\chi$, vanishes for $h_i^{(\rm loc)}\=0$. We indeed observe a similar behavior in the non-equilibrium regime (see also Appendix~\ref{appendix:random_force_fluctuation}). The important point for our purposes here, though, is that $p(\delta{\!f})$ is well-behaved and features a finite standard deviation $\Delta{f}$.

Our main goal then is to understand the relation between $\Delta{f}$ and the stabilized sector --- its characteristic frequency scale in particular --- at the scaling level. To this aim, recall that the stabilized sector is defined through $0\!\le\!k\!\le\!J^2\chi$, whose edge features a characteristic $k\!\simeq\!J^2\chi$ value. Oscillators near the edge of the stabilized sector acquire a characteristic frequency (and displacement $x^*$) from the characteristic force level $\Delta{f}$. That is, at the scaling level, they are described by the following effective potential $v_{\rm eff}(x)\! \simeq\!x^4/4!-\Delta{f}\,x$. The resulting frequency scale, which according to Eq.~\eqref{eq:kappa_w_p_scaling} and Fig.~\ref{fig_monopole_short:Fig5} has to identify with the intrinsic boson peak frequency $\omega_{\rm p}$, satisfies
\begin{equation}
    \omega_{\rm p} \sim (\Delta{f})^{1/3} \ .
\label{eq:w_p_force_relation}
\end{equation}
A similar scaling relation has been suggested in~\cite{Gurevich2003-oo,Parshin2007-od} from a different, complementary perspective.

\begin{figure}[t]
    \centering
    \includegraphics[width=0.95\columnwidth]{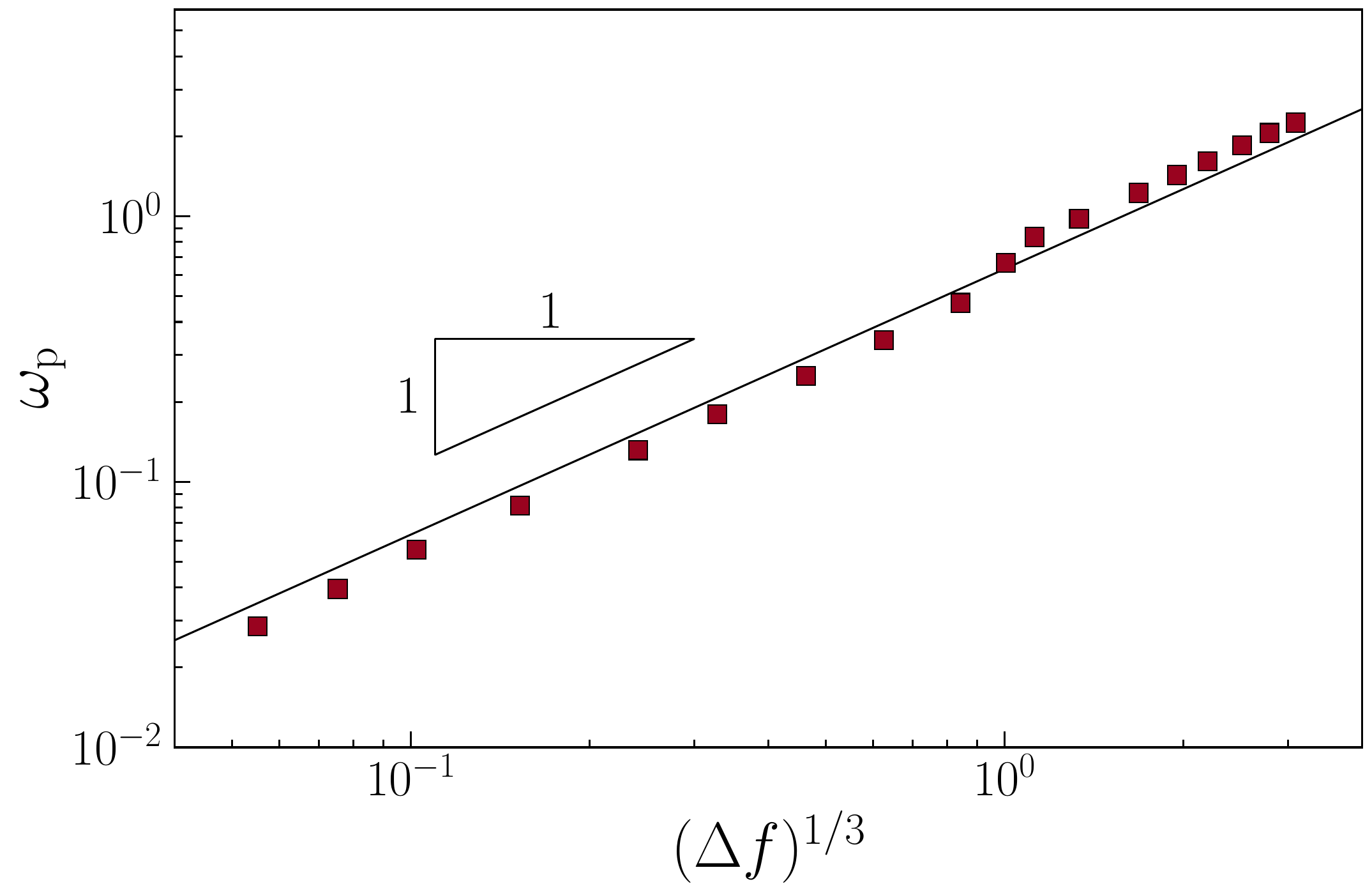}
    \caption{The intrinsic boson-peak frequency $\omega_{\rm p}(J)$ vs.~the characteristic frequency $(\Delta f)^{1/3}$ of the effective potential picture for $N\!=\!8192$ and  $0.01\!\leq\!J\!\leq\!2$ (from bottom-left to top-right), see text for definitions and discussion. The superimposed solid line indicates $\omega_{\rm p}\!\sim\!(\Delta f)^{1/3}$.}
    \label{fig_monopole_short:Fig8}
\end{figure}

In Fig.~\ref{fig_monopole_short:Fig8}, we numerically test Eq.~\eqref{eq:w_p_force_relation} by computing $\Delta{f}(J)$ as detailed above for $0.01\!\leq\!J\!\leq\!2$ and plotting $\omega_{\rm p}(J)$ of Fig.~\ref{fig_monopole_short:Fig5} against $(\Delta{f})^{1/3}$. A favorable agreement is observed over the broad range of quenched disorder levels $J$ used. These results, and the arguments leading to Eq.~\eqref{eq:w_p_force_relation}, further substantiate the importance of the stabilized sector and the central role played by its characteristic frequency/stiffness scale in the model. Overall, this extended correspondence suggests that the effective-potential picture provides a unifying framework for understanding the glassiness of the mean-field model.
%%%%%%%%%%%%%%%%%%%%%%%%%%%%%%%%%%%%%%%%%%%%%%%%%%%%%%%%%%%%
%%%%%%%%%%%%%%%%%%%%%%%%%%%%%%%%%%%%%%%%%%%%%%%%%%%%%%%%%%%%

\section{S\lowercase{ummary and discussion}}

In this paper, we studied a recently introduced mean-field model of glasses, which has already been shown to predict a wide range of glassy phenomena. Our focus was on developing a mean-field micromechanical framework --- based on the force monopole linear response --- and on studying it in two main directions, which turned out to be interrelated. The first aims at testing the predictions of the mean-field model in view of the established corresponding results in finite-dimensional computer glasses. The second aims at gaining a deeper analytic understanding of the model, especially in light of existing numerical evidence that it generically resides in a glassy RSB phase.

In Sect.~\ref{sec:identity}, we derived a mean-field identity that relates the monopole stiffness $\kappa_i$ to other local quantities and to global susceptibilities. The identity shows that the spin-glass susceptibility $\chi_{_{\rm SG}}$ has to be finite in order for $\kappa_i$ to be well defined, which is self-consistently ensured by the universal pseudogapped VDoS tail ${\cal D}_{\!_{\rm G}\!}(\omega)\!\sim\!\omega^4$ of the model, as observed in computer glasses. In Sect.~\ref{sec:VDoS}, we showed that ${\cal D}_{\!_{\rm G}\!}(\omega)$ can be fully recovered over the entire frequency range from $\{\kappa_i\}$ and demonstrated that it is predominantly controlled by a single frequency scale, which determines both the intrinsic boson-peak frequency $\omega_{\rm p}$ and the upper cutoff of the quartic tail, $\omega_{\rm g}$.

In Sect.~\ref{sec:freq_scale}, we used the mean-field identity of Sect.~\ref{sec:identity} to define a stabilization criterion, which selects a subset of the degrees of freedom that are stabilized (or undergo ``reconstruction") when the model's Hamiltonian is quenched to one of its energy landscape minima. We then showed that this stabilized sector reveals a characteristic stiffness/frequency scale, which agrees with the characteristic frequency scale of ${\cal D}_{\!_{\rm G}\!}(\omega)$. In Sect.~\ref{sec:global_vs_mesoscopic}, we further showed that this stiffness scale, which characterizes ``mesoscopic elasticity" in the mean-field model, reveals a larger variation with quenched disorder compared to its global elasticity counterpart, similarly to the corresponding observations in computer glasses.

Finally, we showed in Sect.~\ref{sec:relation_to_effective_potential} that our results --- and the mean-field identity in particular --- indicate the existence of close similarities between the micromechanical monopole response properties of the non-equilibrium RSB phase of the model and an effective potential derived in the equilibrium, zero-temperature RS phase of the model (in some parameter range). These results reinforce the theoretical suggestions of~\cite{Folena2022-tn}. Section~\ref{sec:MF_force_fluctuations} extends this correspondence to include force fluctuations, which are in turn related to the stabilized sector and the boson-peak frequency $\omega_{\rm p}$. Our results yet again indicate that the mean-field model under consideration features a complex energy landscape that is inherently glassy. They also give rise to various future research directions, to be briefly mentioned next.

It would be interesting to explore whether, and if so which of, our results --- obtained using the fully-connected model --- remain valid when the connectivity is reduced, i.e., when the model is defined on a sparse graph~\cite{Baity-Jesi2015-pz,Bernaschi2020-hx,Franz2025-tp}. Likewise, it would be interesting to see whether our results not only reproduce established results in computer glasses, but also offer new predictions, most notably regarding the relations between mesoscopic elasticity and the boson peak in computer and laboratory glasses~\cite{Malinovsky1986-vb,Ramos2022-nf}.

Finally, our results focused on statistical-structural harmonic properties of the model (e.g., its VDoS) and its linear response (derived from the model's Hessian). Future work should be extended to anharmonic properties of the underlying energy landscape and to strongly driven dynamics. In fact, we expect the model's degrees of freedom associated with the stabilized sector to play important roles in its strongly driven elasto-plastic response, already known to reproduce the oscillatory yielding transition, mechanical annealing and memory formation~\cite{Suda2025-kl}.

\vspace{0.5cm}

{\em Acknowledgments} --- E.B.~acknowledges support from the Ben May Center for Chemical Theory and Computation and the Harold Perlman family. M.S.~acknowledges support from
the International Joint Graduate Program in Earth and Environmental Sciences, Tohoku University. The computations were carried out at Weizmann Institute of Science, using the Faculty of Chemistry's high-performance computing facility CHEMFARM, which is supported in part by the Ben May Center for Chemical Theory and Computation.

%\clearpage
\appendix
\setcounter{figure}{0}
\setcounter{equation}{0}
\renewcommand{\thefigure}{A\arabic{figure}}
\renewcommand{\theequation}{\thesection\arabic{equation}}
\setcounter{equation}{0}

\section{T\lowercase{he monopole stiffness mean-field identity and its finite-size fluctuations}}
\label{appendix:mean-field_identity}

Here, we derive and numerically validate the mean-field identity presented in Eq.~\eqref{eq:mean-field_identity}, along with its finite-size fluctuations.
%\subsection{Derivation of the identity and its finite-size fluctuations}
As in the main text, we start by introducing the resolvent $\bm G(z)\!\equiv\!(z\bm I\!-\!\bm{\mathcal{M}})^{-1}$ of the Hessian $\bm{\mathcal{M}}$. For $z\!=\!0$, this resolvent and its derivative yield ${\bm G}(0)\=-\bm{\mathcal{M}}^{-1}$ and ${\bm G}'(0)\=-\bm{\mathcal{M}}^{-2}$. Thus, the monopole stiffness $\kappa_i$ can be expressed in terms of the resolvent diagonals as $\kappa_i\!=\!G_{ii}(0)/G'_{ii}(0)$. We next apply the Schur complement formula~\cite{Bun2017-xj} to the $i^{\rm \tiny th}$ diagonal element of the resolvent, yielding the recursive relation
\begin{equation}
    G_{ii}(z) = \frac{1}{z-\mathcal{M}_{ii}-\sum_{j,k\neq i}\mathcal{M}_{ij}G_{jk}^{(i)}\mathcal{M}}_{ki}\,.
\label{eq:schur_complement}
\end{equation}
Here, ${\bm G}^{(i)}\!\equiv[z\bm I-\bm{\mathcal{M}}^{(i)}]^{-1}$ and $\bm{\mathcal{M}}^{(i)}\!\equiv\!(\mathcal{M}_{jk}\!\!:\!j,k\in[1,N]\backslash \{i\})$ is the minor obtained by removing the $i^{\rm \tiny th}$ row and column of $\bm{\mathcal{M}}$.

Next, we evaluate the quadratic form in the denominator of Eq.~\eqref{eq:schur_complement}. Assuming there is no statistical correlation between the Hessian diagonal $k_i\!+\!\frac{1}{2}x_i^2$ and off-diagonal elements $J_{ij}$~\cite{Bouchbinder2021-dh,Franz2015-cc}, and using the Gaussian statistics of $J_{ij}$, one obtains
\begin{equation}
    \sum_{j,k\neq i}\!\mathcal{M}_{ij}G_{jk}^{(i)}\mathcal{M}_{ki} = \frac{J^2}{N}\sum_{j\neq i}G_{jj}^{(i)}+\mathcal{O}\!\left(\!\sqrt{{\rm Tr}({\bm G}^{(i)})^2/N^2}\right)\,.
\label{eq:schur_quadratic}
\end{equation}
The last term on the RHS corresponds to finite-size fluctuations, which is the standard deviation of the same quadratic form.

Setting $z\!=\!0$, and using $G_{jj}^{(i)}\!=\!G_{jj}+\mathcal{O}(N^{-1})$ and $\chi_{_{\rm SG}}\!=\!N^{-1}\,{\rm Tr}[\bm G(0)]^2$, the recursive relation in Eq.~\eqref{eq:schur_complement} becomes
\begin{equation}
    G_{ii}(0)=-\frac{1}{(k_i-J^2\chi)+\frac{1}{2}x_i^2}+\mathcal{O}\!\left(\!\sqrt{\frac{\chi_{_{\rm SG}}}{N}}\,\right)\,,
\label{eq:G_self-consistency}
\end{equation}
once combined with Eq.~\eqref{eq:schur_quadratic}. Repeating the same procedure for the derivative $G'_{ii}(0)$ gives
\begin{equation}
    G_{ii}'(0)=-\frac{1+J^2\chi_{_{\rm SG}}}{[(k_i-J^2\chi)+\frac{1}{2}x_i^2]^2}+\mathcal{O}\!\left(\!\sqrt{\frac{\chi^{(4)}}{N}}\,\right)\,,
\label{eq:G'_self-consistency}
\end{equation}
where $\chi^{(4)}\!\equiv\!N^{-1}\,{\rm Tr}(\bm{\mathcal{M}}^{-4})$ is the $4^{\rm \tiny th}$ order susceptibility of $\bm{\mathcal{M}}$. By taking the ratio $\kappa_i\=G_{ii}(0)/G'_{ii}(0)$, one obtains Eq.~\eqref{eq:mean-field_identity} to leading order in $N$, but also the finite-$N$ fluctuations/corrections, which we consider next.

The finite-$N$ contributions to $\kappa_i\=G_{ii}(0)/G'_{ii}(0)$, in view of Eqs.~\eqref{eq:G_self-consistency}--\eqref{eq:G'_self-consistency}, involve $\sqrt{\chi_{_{\rm SG}}/N}$ and $\sqrt{\chi^{(4)}/N}$ additively. Consequently, we need to determine the $N$ dependence of these two quantities. As noted in Sect.~\ref{sec:identity}, the $N$ dependence of the global susceptibilities $\chi_{_{\rm SG}}\!\sim\!\int\!\omega^{-4}\,{\cal D}_{\!_{\rm G}\!}(\omega)\,d\omega$ and $\chi^{(4)}\!\sim\!\int\!\omega^{-8}\,{\cal D}_{\!_{\rm G}\!}(\omega)\,d\omega$ is inherited from the finite-$N$ scaling of the minimal frequency $\omega_{\rm min}$. Taking into account the universal quartic tail ${\cal D}_{\!_{\rm G}\!}(\omega)\!\sim\!\omega^4$, the scaling of the the mean of the minimal frequency, $\langle\omega_{\rm min}\rangle$, follows from a standard extreme-value argument that reads
\begin{equation}
    \int_0^{\langle\omega_{\rm min}\rangle}\!\!{\cal D}_{\!_{\rm G}\!}(\omega)\,d\omega \sim N^{-1}\quad\Longrightarrow\quad \langle\omega_{\rm min}\rangle \sim N^{-1/5}\,.
\end{equation}

Combining all the above, we obtain $\chi^{(4)}\!\sim\!\int_{_{\langle\omega_{\rm min}\rangle}}\!\omega^{-8}\,{\cal D}_{\!_{\rm G}\!}(\omega)\,d\omega\!\sim\!N^{3/5}$, from which we obtain $\sqrt{\chi^{(4)}/N}\!\sim\!N^{-1/5}$. Repeating this scaling analysis for $\sqrt{\chi_{_{\rm SG}}/N}$, and recalling that the two finite-$N$ contributions enter $\kappa_i\=G_{ii}(0)/G'_{ii}(0)$ additively, we find that $\sqrt{\chi_{_{\rm SG}}/N}\!\ll\!\sqrt{\chi^{(4)}/N}\!\sim\!N^{-1/5}$ is the dominant contribution. We then define
\begin{equation}
    \Delta_{\kappa_i} \equiv \kappa_i-\frac{(k_i-J^2\chi)+\frac{1}{2}x_i^2}{1+J^2\chi_{_{\rm SG}}}\,,
\label{eq:Delta_kappa}
\end{equation}
where $\kappa_i$ is understood to be the directly measured monopole stiffness (cf.~Eq.~\eqref{eq:monopole_stiffness}) in a finite-$N$ realization of the model. That is, $\Delta_{\kappa_i}$ is the finite-$N$ difference between the measured $\kappa_i$ and the analytic prediction on the RHS of Eq.~\eqref{eq:mean-field_identity}, which is exact for $N\!\to\infty$. The above analysis suggests that $p(\Delta_{\kappa_i})$ would feature
\begin{equation}
    \langle \Delta_{\kappa_i} \rangle \ll \sqrt{\mu'_2(\Delta_{\kappa_i})} \sim N^{-1/5} \ ,
\label{eq:2nd_moment}
\end{equation}
where $\mu'_2(\Delta_{\kappa_i})$ is the second moment of $p(\Delta_{\kappa_i})$.

\begin{figure}[t]
    \centering
    \includegraphics[width=0.95\columnwidth]{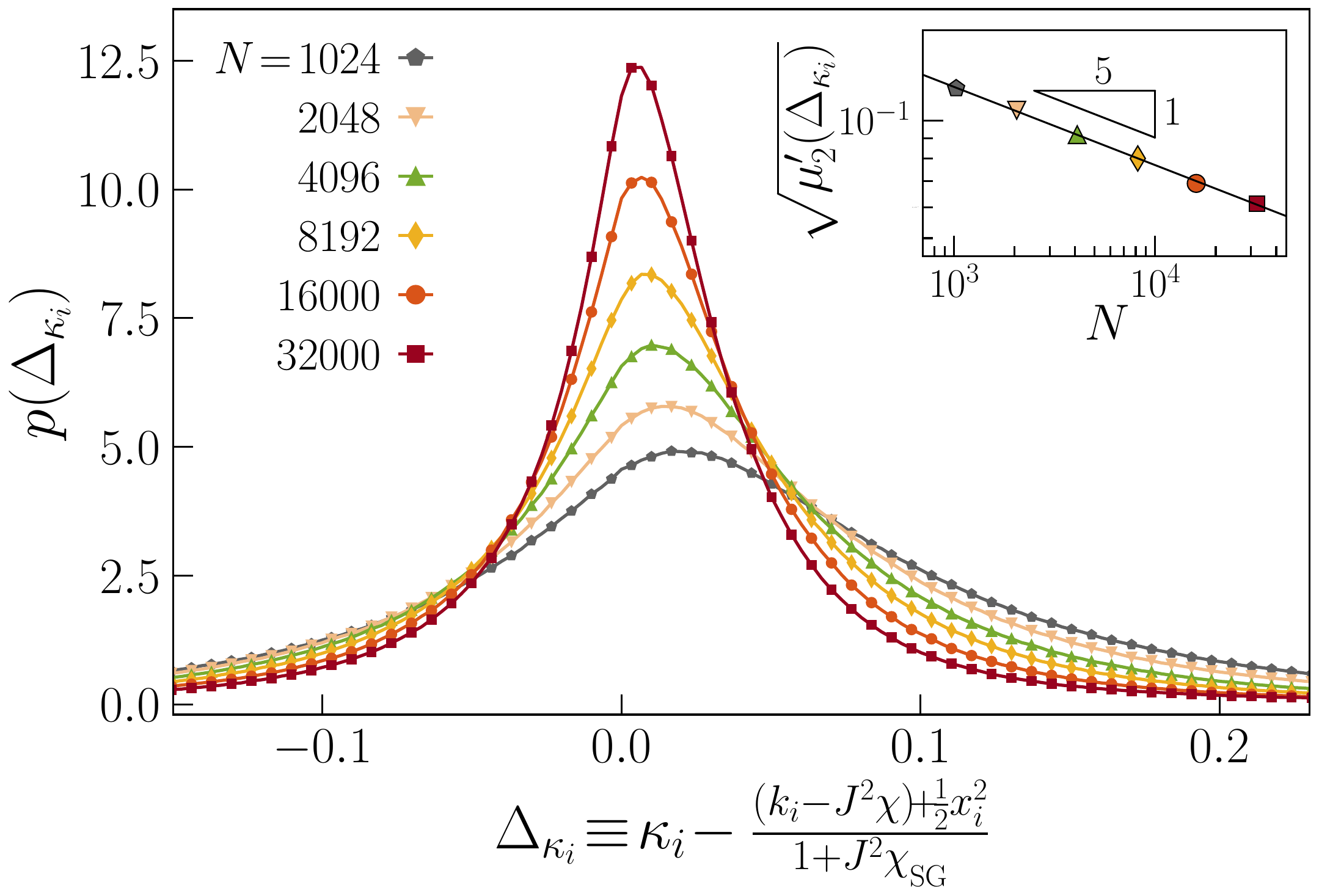}
    \caption{The distribution $p(\Delta_{\kappa_i})$ of the error between the measured monopole stiffness $\kappa_i$ and its analytic prediction in Eq.~\eqref{eq:mean-field_identity} at $J\!=\!0.5$ and various $N$ (see legend). (inset) The square root of the second moment $\sqrt{\mu_2'(\Delta_{\kappa_i})}$ vs. the system size $N$. The superimposed solid line indicates $\sqrt{\mu_2'(\Delta_{\kappa_i})}\!\sim\!N^{-1/5}$. See text for discussion.}
    \label{fig_monopole_short:FigS1}
\end{figure}

The prediction in Eq.~\eqref{eq:2nd_moment} reveals anomalous fluctuations, which slowly decay with $N$. Our next goal is to numerically test this prediction, which amounts to testing the mean-field identity (recall the definition in Eq.~\eqref{eq:Delta_kappa}) and its finite-size fluctuations. The results are presented in Fig.~\ref{fig_monopole_short:FigS1}, demonstrating that $p(\Delta_{\kappa_i})$ is centered around zero and it becomes progressively narrower with increasing $N$. The latter is quantified in the inset, confirming the predicted $\sqrt{\mu'_2(\Delta_{\kappa_i})}\!\sim\!N^{-1/5}$ scaling. These findings provide direct support to the validity of our main result in Eq.~\eqref{eq:mean-field_identity} and its convergence with increasing $N$.

%%%%%%%%%%%%%%%%%%%%%%%%%%%%%%%%%%%%%%%%%%%%%%%%%%%%%%%%%%%%

\section{C\lowercase{orollaries of the mean-field identity}}
\label{appendix:related_identities}

\subsection{The pair identities}

The mean-field identity in Eq.~\eqref{eq:mean-field_identity} can be also used to construct pairwise, \textit{local} estimators of the global susceptibilities $\chi$ and $\chi_{_{\rm SG}}$. Applying the mean-field identity in Eq.~\eqref{eq:mean-field_identity} to a pair of oscillators, say the $i^{\tiny \rm th}$ and $j^{\tiny \rm th}$ ones, and solving the resulting two equations for the unknown susceptibilities, one obtains
\begin{equation}
    J^2\chi^{(ij)}\!=\!\frac{a_j\kappa_i\!-\!a_i\kappa_j}{\kappa_i\!-\!\kappa_j}\,,\qquad J^2\chi_{_{\rm SG}}^{(ij)}\=\frac{a_i\!-\!a_j}{\kappa_i\!-\!\kappa_j}\!-\!1\,,
\end{equation}
where $a_i\!\equiv\!k_i\!+\!\frac{1}{2}x_i^2$ and the superscript $(ij)$ indicates that these susceptibilities are estimated from a single pair of oscillators.

We test these estimates against the values obtained directly from the inverse moments of the Hessian, $\chi\!=\!N^{-1}\,{\rm Tr}(\bm{\mathcal{M}}^{-1})$ and $\chi_{_{\rm SG}}\!=\!N^{-1}\,{\rm Tr}(\bm{\mathcal{M}}^{-2})$. We define the errors
\begin{equation}
    \Delta_{\chi}^{(ij)}\!\equiv\!\chi^{(ij)}\!-\!\chi\,,\qquad
    \Delta_{\chi_{\rm SG}}^{(ij)}\!\equiv\!\chi^{(ij)}_{_{\rm SG}}\!-\!\chi_{_{\rm SG}}\,,
\end{equation}
and analyze their distributions, $p(\Delta_{\chi}^{(ij)})$ and $p(\Delta_{\chi_{\rm SG}}^{(ij)})$, and the associated second moments, $\mu'_2(\Delta^{(ij)}_\chi)$ and $\mu'_2(\Delta^{(ij)}_{\chi_{\rm SG}})$. Since these errors inherit the leading finite-size fluctuations of the main mean-field identity, we expect $\sqrt{\mu'_2(\Delta^{(ij)}_\chi)},\,\sqrt{\mu'_2(\Delta^{(ij)}_{\chi_{\rm SG}})}\!\sim\!N^{-1/5}$.

\begin{figure}[t]
    \centering
    \includegraphics[width=0.95\columnwidth]{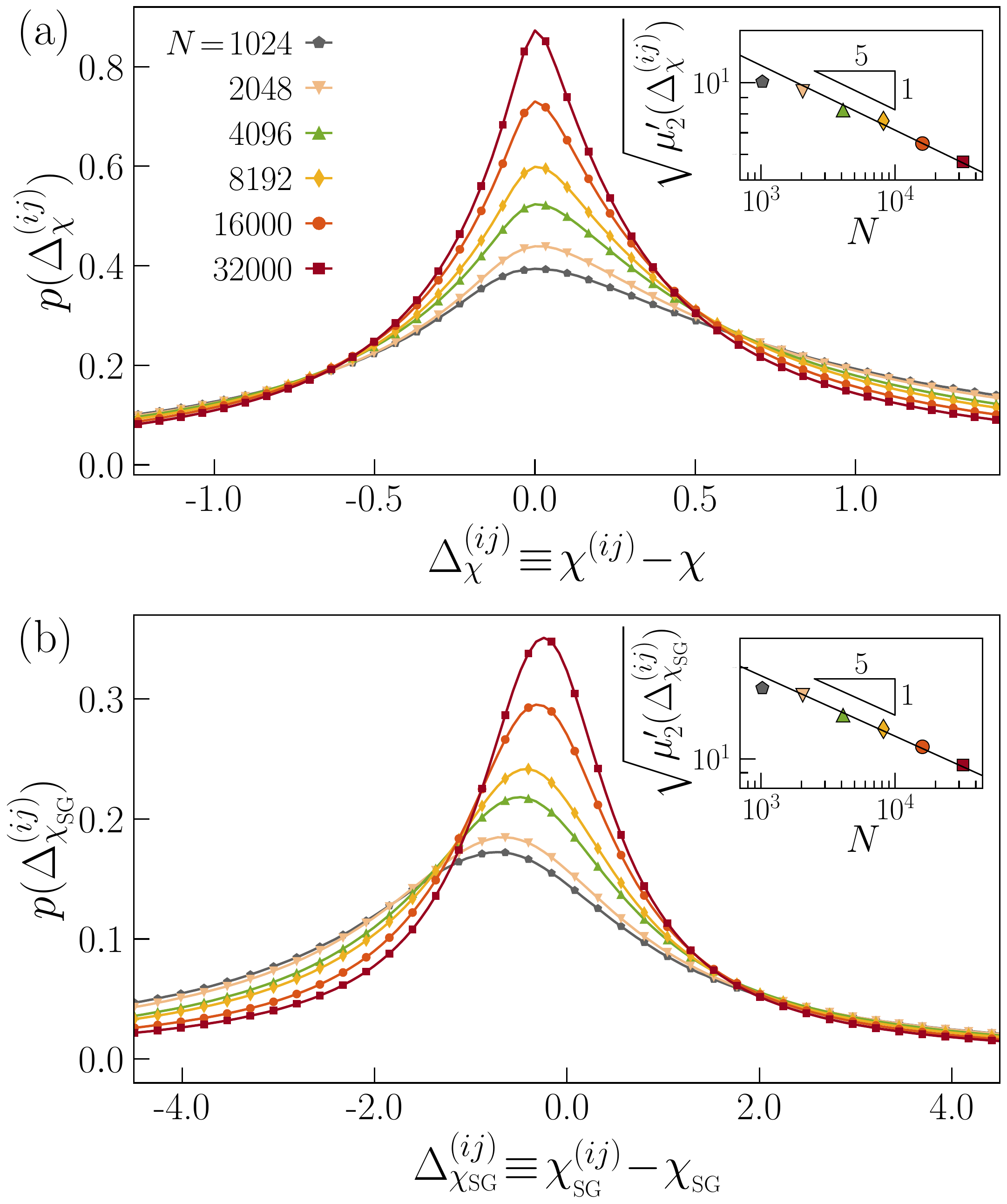}
    \caption{(a) The distribution $p(\Delta_{\chi}^{(ij)})$ of the error associated with the pairwise estimator of the linear susceptibility $\chi$ (inset) The square root of the second moment $\sqrt{\mu_2'(\Delta_\chi^{(ij)})}$ vs. the system size $N$. The superimposed solid line indicates $\sqrt{\mu_2'(\Delta_\chi^{(ij)})}\!\sim\!N^{-1/5}$. (b) Same as (a), but for the error $\Delta_{\chi_{\rm SG}}^{(ij)}$ associated with the pairwise estimator of the spin-glass susceptibility $\chi_{\rm SG}$. (inset) Same as (a, inset), but for $\Delta_{\chi_{\rm SG}}^{(ij)}$. Here, we use $J\!=\!0.5$ and various $N$ (see legend). See text for definitions and discussion.}
    \label{fig_monopole_short:FigS2}
\end{figure}

Figure~\ref{fig_monopole_short:FigS2} shows the distributions $p(\Delta_{\chi}^{(ij)})$ and $p(\Delta_{\chi_{\rm SG}}^{(ij)})$. Both distributions are peaked near zero, and both narrow systematically as $N$ increases. To avoid artificial amplification of the errors, we compute the second moments only from the oscillator pairs satisfying $|\kappa_i\!-\!\kappa_j|\!>\!10^{-2}$. The insets show that $\sqrt{\mu'_2(\Delta^{(ij)}_\chi)}$ and $\sqrt{\mu'_2(\Delta^{(ij)}_{\chi_{\rm SG}})}$ follow the predicted $N^{-1/5}$ scaling. Consequently, in the limit $N\!\to\!\infty$, the global susceptibilities $\chi$ and $\chi_{_{\rm SG}}$ can be recovered from purely local measurements on oscillator pairs.

%%%%%%%%%%%%%%%%%%%%%%%%%%%%%%%%%%%%%%%%%%%%%%%%%%%%%%%%%%%%
%%%%%%%%%%%%%%%%%%%%%%%%%%%%%%%%%%%%%%%%%%%%%%%%%%%%%%%%%%%%

\subsection{The global identities}

\begin{figure}[t]
    \centering
    \includegraphics[width=0.95\columnwidth]{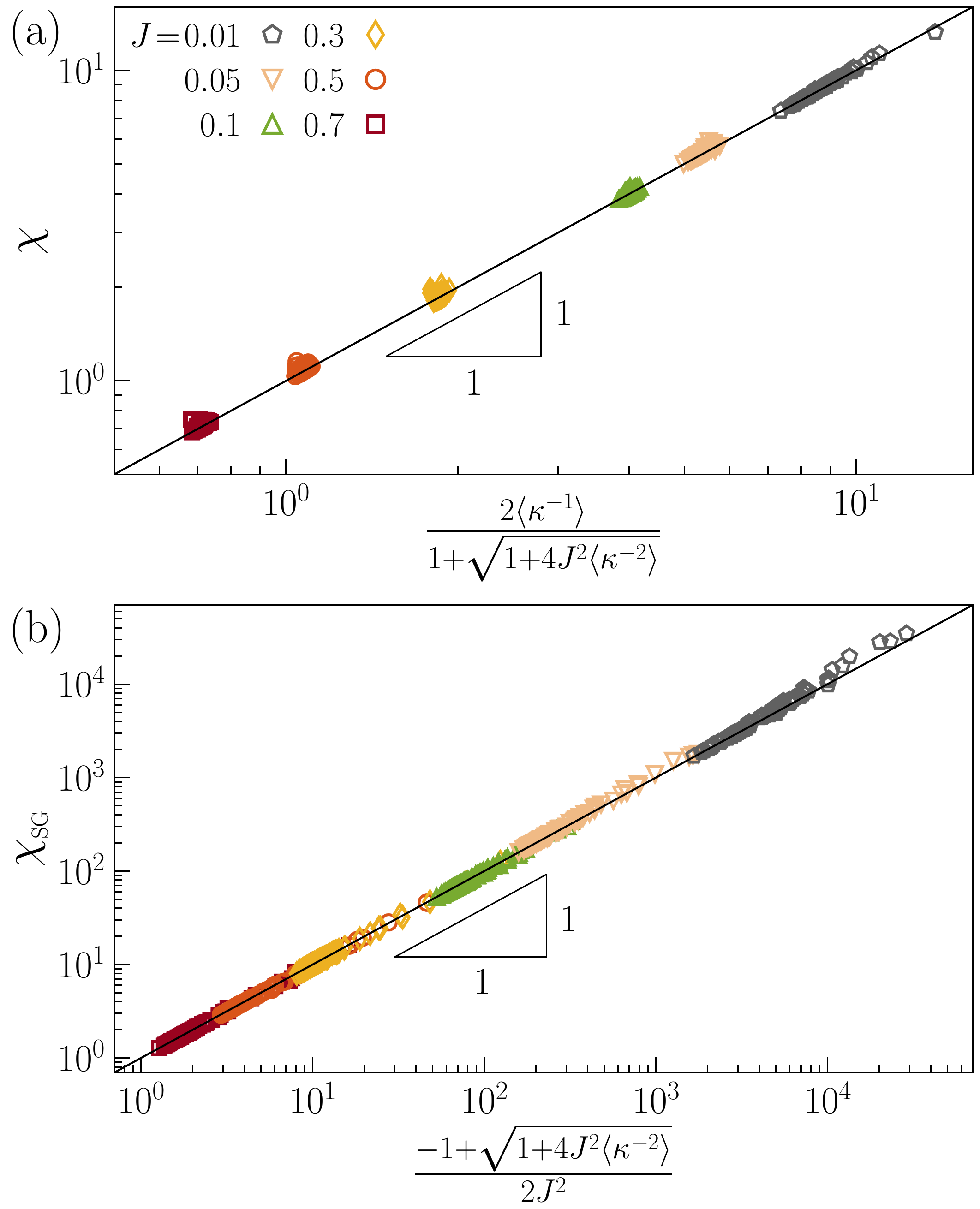}
    \caption{(a) The linear susceptibility $\chi\!\equiv\!N^{-1}\,{\rm Tr}(\bm{\mathcal{M}}^{-1})$ vs. its analytic prediction in Eqs.~\eqref{eq:global_identities} for various $J$ values and $N\!=\!8192$ (see legend). Each data point corresponds to a single disorder realization and herein $M\!=\!200$ realizations are plotted. The superimposed solid line corresponds to the identity. (b) Same as (a), but for the spin-glass susceptibility $\chi_{_{\rm SG}}\!\equiv\!N^{-1}\,{\rm Tr}(\bm{\mathcal{M}}^{-2})$. See text for definitions and discussion.}
    \label{fig_monopole_short:FigS3}
\end{figure}

Yet another consequence of the mean-field identity is a pair of \textit{global} identities relating the inverse moments of the monopole stiffness $\kappa_i$ to the susceptibilities $\chi$ and $\chi_{_{\rm SG}}$. The first step is to average the self-consistency Eqs.~\eqref{eq:G_self-consistency}--\eqref{eq:G'_self-consistency} over all oscillators. Neglecting the finite-size fluctuations, one obtains
\begin{equation}
    \chi=\left\langle\frac{1}{a-J^2\chi}\right\rangle\,,\qquad
    \chi_{_{\rm SG}} = \left\langle\frac{1+J^2\chi_{_{\rm SG}}}{(a-J^2\chi)^2}\right\rangle\,,
\end{equation}
where $\langle\cdot\rangle$ denotes the average over $N$ oscillators (and hence we dropped the subscript $i$). Combining these relations with the mean-field identity in Eq.~\eqref{eq:mean-field_identity} gives the self-consistency relations for the inverse moments $\langle\kappa^{-1}\rangle$ and $\langle\kappa^{-2}\rangle$,
\begin{equation}
\begin{split}
    \hspace{-0.2cm}\langle\kappa^{-1}\rangle \!=\! (1\!+\!J^2\chi_{_{\rm SG}})\chi\,,\qquad
    \langle\kappa^{-2}\rangle \!=\! (1\!+\!J^2\chi_{_{\rm SG}})\chi_{_{\rm SG}}\,.
\end{split}
\end{equation}
Solving these relations in favor of $\chi$ and $\chi_{_{\rm SG}}$, one obtains
\begin{equation}
    \chi\!=\!\frac{2\langle\kappa^{-1}\rangle}{1\!+\!\sqrt{1\!+\!4J^2\langle\kappa^{-2}\rangle}}\,,\qquad \chi_{_{\rm SG}}\!=\!\frac{-1\!+\!\sqrt{1\!+\!4J^2\langle\kappa^{-2}\rangle}}{2J^2}\,.
\label{eq:global_identities}
\end{equation}
Equations~\eqref{eq:global_identities} express the global susceptibilities entirely in terms of the inverse moments $\langle\kappa^{-1}\rangle$ and $\langle\kappa^{-2}\rangle$, and disorder level $J$, without an explicit reference to the Hessian diagonal elements $a_i$. In Fig.~\ref{fig_monopole_short:FigS3}, we numerically test Eqs.~\eqref{eq:global_identities} for $0.01\!\le\!J\!\le\!0.7$, demonstrating their validity. Overall, these results strongly support the derived identities and confirm that, in the framework of the present mean-field model, the global susceptibilities $\chi$ and $\chi_{_{\rm SG}}$ can be obtained from the statistics of monopole stiffness alone.
%%%%%%%%%%%%%%%%%%%%%%%%%%%%%%%%%%%%%%%%%%%%%%%%%%%%%%%%%%%%
%%%%%%%%%%%%%%%%%%%%%%%%%%%%%%%%%%%%%%%%%%%%%%%%%%%%%%%%%%%%

\section{D\lowercase{etails of the} VD\lowercase{o}S \lowercase{reconstruction from monopole responses}}
\label{appendix:VDoS_reconstruction}

% In Sect.~\ref{sec:VDoS}, we showed that the model's VDoS can be reconstructed from monopole responses alone. The central quantity in obtaining this result is

In Sect.~\ref{sec:VDoS}, we showed that the model's VDoS can be reconstructed from monopole responses alone. Here, we describe the procedure in detail. The central quantity underlying this result is the normalized trace of the resolvent $\mathfrak{g}_{_{\bm{\mathcal{M}}}}\!(z)\!\equiv\!N^{-1}\,{\rm Tr}[\bm{G}(z)]\!=\!N^{-1}\,{\rm Tr}[(z\bm I\!-\!\bm{\mathcal{M}})^{-1}]$. From this quantity, the eigenvalue $\lambda\!=\!\omega^2$ density is calculated as~\cite{Bun2017-xj}
\begin{equation}
    \rho_{_{\bm{\mathcal{M}}}}(\lambda)=\frac{1}{\pi}\lim_{\eta\to0^{+}}\Im\,{[\mathfrak{g}_{_{\bm{\mathcal{M}}}}\!(\lambda-i\eta)]}\,,
\label{eq:resolvent_limit}
\end{equation}
and the VDoS is obtained through $\mathcal{D}_{\!_{\rm G}\!}(\omega)\!=\!2\omega\rho_{_{\bm{\mathcal{M}}}}\!(\omega^2)$. In the present mean-field model, evaluating the self-consistency relation in Eq.~\eqref{eq:G_self-consistency} at complex $z$ and taking the average over $N$ oscillators on both sides, one obtains Eq.~\eqref{eq:resolvent_self-consistency},
% \begin{equation}
%     \mathfrak{g}_{_{_{\bm{\mathcal{M}}}}}\!(z)=N^{-1}\sum_i\frac{1}{z-a_i-J^2\,\mathfrak{g}_{_{_{\bm{\mathcal{M}}}}}\!(z)}\,,
% \label{eq:resolvent_self_consistency}
% \end{equation}
which self-consistently determines $\mathfrak{g}_{_{_{\bm{\mathcal{M}}}}}\!(z)$ once the set $\{a_i\}$ is known.

Our idea is to obtain $\{a_i\}$ from $\{\kappa_i\}$, which can be done by combining Eqs.~\eqref{eq:global_identities} with Eq.~\eqref{eq:mean-field_identity}, yielding
\begin{equation}
    a_i=\frac{1\!+\!\sqrt{1\!+\!4J^2\langle\kappa^{-2}\rangle}}{2}\kappa_i+\frac{2J^2\langle\kappa^{-1}\rangle}{1\!+\!\sqrt{1\!+\!4J^2\langle\kappa^{-2}\rangle}}\,.
\label{eq:a_i_identity}
\end{equation}
Equation~\eqref{eq:a_i_identity} shows that the statistics of the monopole stiffness $\{\kappa_i\}$, from which the inverse moments $\langle\kappa^{-1}\rangle$ and $\langle\kappa^{-2}\rangle$ can be extracted, allow one to obtain the statistics of $\{a_i\}$.

The above results delineate the following procedure; first, the monopole stiffnesses $\{\kappa_i\}$ are collected, from which $\langle\kappa^{-1}\rangle$ and $\langle\kappa^{-2}\rangle$ are computed, leading to $\{a_i\}$ based on Eq.~\eqref{eq:a_i_identity}. Second, $\{a_i\}$ are used to solve the resolvent self-consistency in Eq.~\eqref{eq:resolvent_self-consistency} in favor of the continuous complex-variable function $\mathfrak{g}_{_{\bm{\mathcal{M}}}}\!(z)$. Finally, Eq.~\eqref{eq:resolvent_limit} can be used to obtain $\mathcal{D}_{\!_{\rm G}\!}(\omega)$. As described in Sect.~\ref{sec:VDoS}, this procedure amounts to the following sequence of transformations: $p(\kappa)\!\to\!p(a)\!\to\!\mathfrak{g}_{_{_{\bm{\mathcal{M}}}}}\!(z)\!\to\!{\cal D}_{\!_{\rm G}\!}(\omega)$.

In practice, we consider the model with $N\!=\!8192$ and for each disorder realization at a fixed $J$ we compute all monopole stiffnesses $\{\kappa_i\}_{i=1}^{i=N}$ following its original definition in Eq.~\eqref{eq:monopole_stiffness}. These stiffnesses are then transformed into $\{a_i\}_{i=1}^{i=N}$ through Eq.~\eqref{eq:a_i_identity}. Then, Eq.~\eqref{eq:resolvent_self-consistency} is solved iteratively for $\mathfrak{g}_{_{\bm{\mathcal{M}}}}\!(z;N)$ over a discrete grid $z_{j}\!\equiv\!\lambda_j-i\eta$ with a real $\lambda_j$ and a very small $\eta\!=\!10^{-8}$ until convergence. Once achieved, $\rho_{_{\bm{\mathcal{M}}}}\!(\lambda_j;N)$ is obtained, from which the VDoS is estimated for the specific realization. This procedure is repeated for $M\!=\!2000$ independent disorder realizations and $\mathcal{D}_{\!_{\rm G}\!}(\omega)$ is obtained by averaging over the $M$ realizations per $J$. The resulting $\mathcal{D}_{\!_{\rm G}\!}(\omega)$ is presented in Fig.~\ref{fig_monopole_short:Fig3}a.

\section{E\lowercase{xtracting the cutoff frequency $\omega_{\rm g}$}}
\label{appendix:omega_g}

In Fig.~\ref{fig_monopole_short:Fig3}b, we estimated the upper cutoff frequency $\omega_{\rm g}(J)$ of the quartic tail of $\mathcal{D}_{\!_{\rm G}\!}(\omega)$. The procedure involves two steps. First, we fit $\mathcal{D}_{\!_{\rm G}\!}(\omega)\!=\!A_{\rm g}(J)\,\omega^4$ to the low-frequency tail of the VDoS at each $J$. The VDoS is computed for $N\!=\!8192$ and $M\!=\!2000$ disorder realizations. Second, we operationally define $\omega_{\rm g}(J)$ as the smallest frequency satisfying
$A_{\rm g}(J)\,\omega^4/\mathcal{D}_{\!_{\rm G}\!}(\omega)\!\geq\!2$. Before evaluating this ratio, we apply a cubic interpolation to the numerical VDoS to improve the precision of extracted $\omega_{\rm g}(J)$. This procedure is illustrated in the inset of Fig.~\ref{fig_monopole_short:Fig3}b.

\section{T\lowercase{he global modulus }$G$}
\label{appendix:global_modulus}

In Sect.~\ref{sec:global_vs_mesoscopic}, we defined a global elastic modulus (stiffness) $G$ in the mean-field model, in analogy to linear elastic moduli in finite-dimensional solids, and provided an expression for it in Eq.~\eqref{eq:modulus}. Here, we briefly present a derivation of the latter, starting from the definition $G^{-1}\!\equiv\!d^2(H/N)/df^2$ given therein.

Using the Hamiltonian $H$ in Eq.~\eqref{eq:Hamiltonian} and the above definition, we obtain $G^{-1}\=N^{-1}\,\bm 1\cdot d\bm x/df$, where $\bm 1\!\equiv\!(1,\dots,1)$. As $d\bm x/df$ is evaluated at a minimum of $H$, differentiating the mechanical equilibrium condition, $\pa H/\pa \bm x\=\bm 0$, with respect to $f$ gives
\begin{equation}
    \bm 0 = \frac{d}{df}\frac{\pa H}{\pa \bm x} = \frac{\pa^2 H}{\pa \bm x\pa \bm x}\cdot\frac{d\bm x}{df}-\bm 1=\bm{\mathcal{M}}\cdot \frac{d\bm x}{df}-\bm 1\,.
\end{equation}
Consequently
\begin{equation}
    \frac{d\bm x}{df} = \bm{\mathcal{M}}^{-1}\cdot \bm 1\,,
\end{equation}
from which Eq.~\eqref{eq:modulus} follows.

% \section{M\lowercase{arginality of the sampled minima}}
% \label{appendix:marginality}

% $$\sum_i h_i\,x_i$$

% $$\langle\tilde h_i^2\rangle/\langle h_i^2\rangle$$

% Consider an infinitesimal random field $\bm h$ with $\langle h_i\rangle\!=\!0$ and $\langle h_ih_j\rangle\!=\!h^2\delta_{ij}$. Within linear response, the resulting displacement is $\delta \bm x\!=\bm{\mathcal{M}}^{-1}\cdot\bm h$, and this displacement in turn generates an interaction-induced field $\tilde{\bm h}$ where $\tilde h_i\!=\!\sum_j J_{ij}\delta x_j$. Evaluating the variance of the induced field, one obtains
% \begin{equation}
%     \left\langle\tilde h_i^2\right\rangle\!=\!\frac{J^2}{N}\sum_j\langle\delta x_j^2\rangle \!=\! \frac{h^2J^2}{N}{\rm Tr}(\bm{\mathcal{M}}^{-2})\!=\!h^2J^2\chi_{_{\rm SG}}\,.
% \end{equation}
% Thus, $J^2\chi_{_{\rm SG}}$ measures the gain in perturbation variance after one self-feedback loop. The numerically observed condition $J^2\chi_{_{\rm SG}}\!=\!1$ therefore implies that the sampled minima are marginal with respect to random-field perturbations.

\section{T\lowercase{he random force distribution and its characteristic scale}}
\label{appendix:random_force_fluctuation}

\begin{figure}[t!]
    \centering
    \includegraphics[width=0.95\columnwidth]{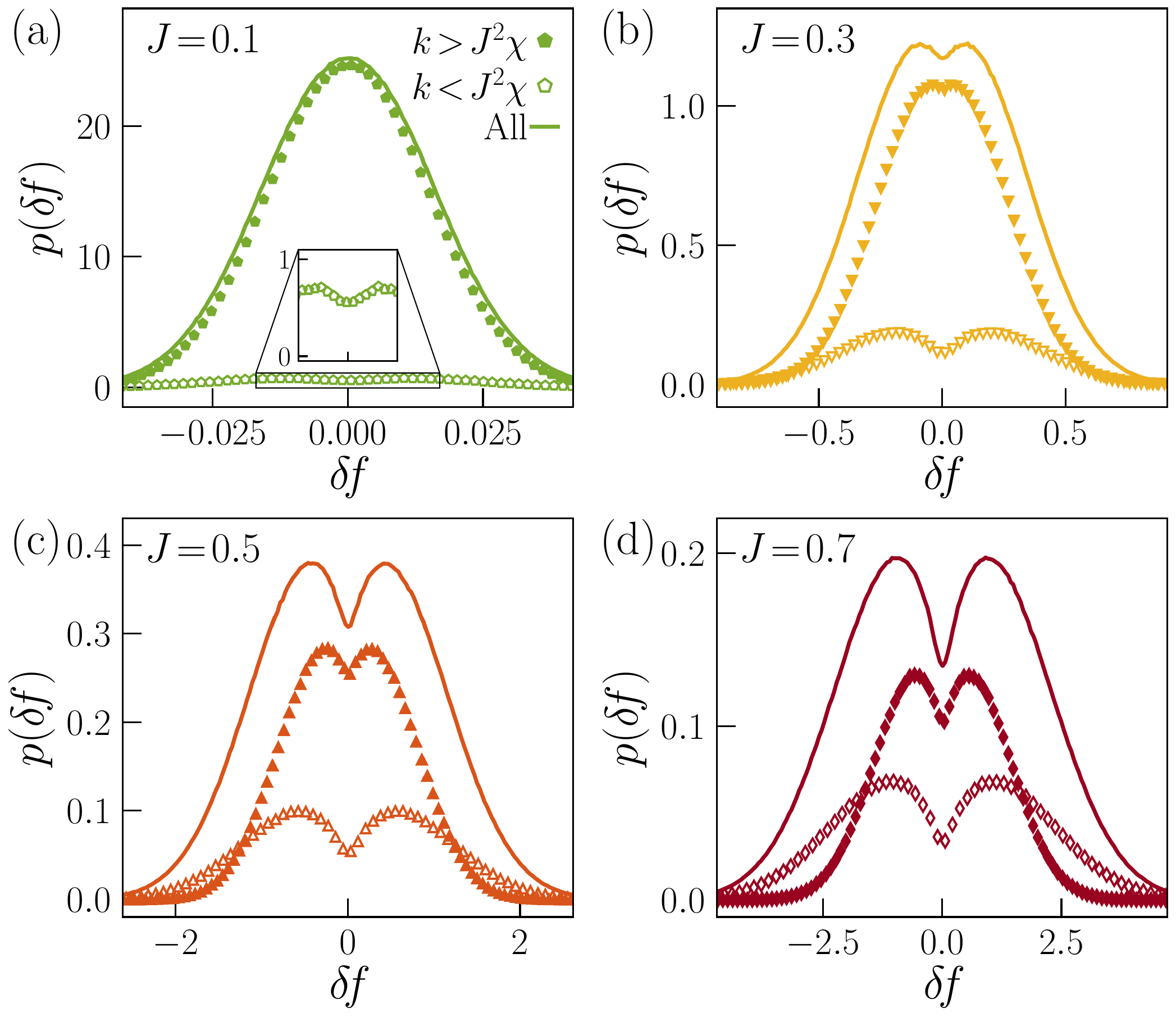}
    \caption{The distributions $p(\delta{\!f})$ of random force fluctuations, along with its associated conditional distributions in the stabilized $k\!<\!J^2\chi$ and the complementary $k\!>\!J^2\chi$ sectors, for various $J$ values at $N\!=\!8192$ (see legend). In panel (a), we also present a zoom in on the small $|\delta\!f|$ portion of the stabilized-sector distribution. See text for definitions and discussion.}
    \label{fig_monopole_short:FigS4}
\end{figure}

The random force $\delta{\!f}_{\!i}$ is discussed in Sect.~\ref{sec:MF_force_fluctuations}. Here, we provide additional details about the associated calculations and some complementary results. We extract $\delta{\!f}_{\!i}$ in two steps. For each sampled minimum of the Hamiltonian in Eq.~\eqref{eq:Hamiltonian}, we first compute $\chi\!\equiv\!N^{-1}\,{\rm Tr}(\bm{\mathcal{M}}^{-1})$, and then use the mapping $\delta{\!f}_{\!i}\!=\!(k_i-J^2\chi)x_i+x_i^3/6$ implied by mechanical equilibrium in the effective potential. Repeating this process over $M\!=\!2000$ disorder realizations yields the disorder-averaged $p(\delta{\!f})$ and its standard deviation $\Delta f$ at each $J$. Figure~\ref{fig_monopole_short:FigS4} shows the distribution $p(\delta{\!f})$, along with its associated conditional distributions in the stabilized and its complementary sectors. As required by the symmetry, $p(\delta{\!f})$ is symmetric about the origin and satisfies $\langle\delta{\!f}\rangle\!=\!0$. $p(\delta{\!f})$ is approximately Gaussian for $J\!\lesssim\!0.3$ and becomes bimodal at larger $J$, while retaining exponentially decaying tails that allow its standard deviation $\Delta{f}$ to be extracted robustly.

\begin{figure}[t]
    \centering
    \includegraphics[width=0.95\columnwidth]{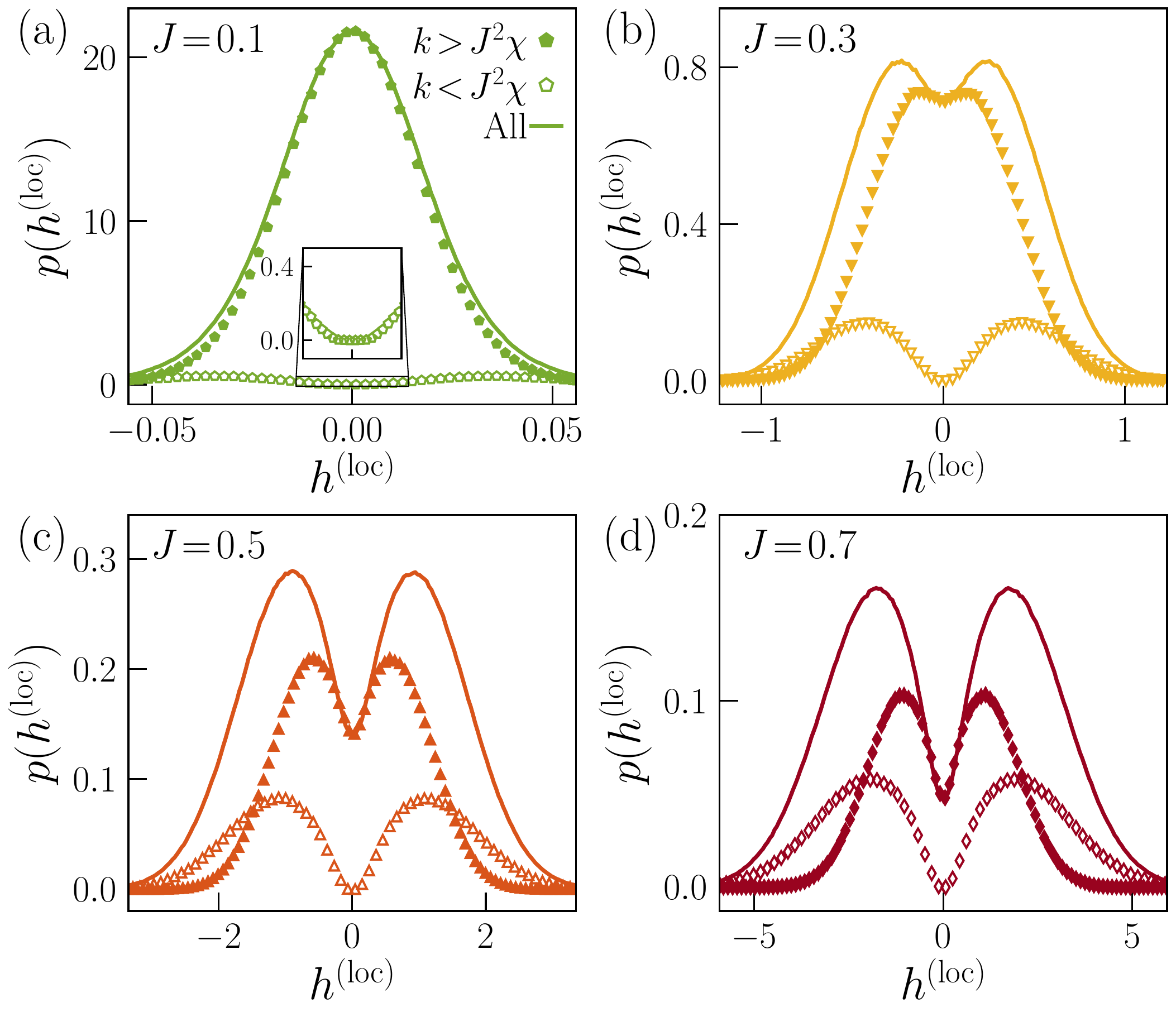}
    \caption{The same as Fig.~\ref{fig_monopole_short:FigS4} but for $h_i^{(\rm loc)}\!=\!\delta\!f_i\!+\!J^2\chi\,x_i$, see text for discussion.}
    % \caption{The distributions $p(h^{\rm (loc)})$, along with its associated conditional distributions in the stabilized and the complementary sectors, for various $J$ values at $N\!=\!8192$ (see legend). In panel (a), we also present a zoom in on the small $|h^{\rm (loc)}|$ portion of the stabilized-sector distribution. See text for definitions and discussion.}
    \label{fig_monopole_short:FigS5}
\end{figure}

In fact, $p(\delta{\!f}|k\!<\!J^2\chi)$ shown in Fig.~\ref{fig_monopole_short:FigS4} features a dip near the origin for all $J$ values. This observation appears to be connected to a claim made in~\cite{Folena2022-tn} about a related quantity, defined as $h_i^{(\rm loc)}\!\equiv\!\delta{\!f}_{\!i}+J^2\chi\,x_i$. It is suggested therein that
right beyond the RSB transition, $p(h^{(\rm loc)}|k\!<\!J^2\chi)$ vanishes for $h^{(\rm loc)}\=0$. Indeed, plotting $p(h^{(\rm loc)}|k\!<\!J^2\chi)$ in Fig.~\ref{fig_monopole_short:FigS5} using the same data as in Fig.~\ref{fig_monopole_short:FigS4}, reveals a similar behavior. This similarity further strengthens the effective potential description of the model's non-equilibrium behavior.

% while retaining sufficiently regular tails for its standard deviation $\Delta f$ to be extracted robustly.
%%%%%%%%%%%%%%%%%%%%%%%%%%%%%%%%%%%%%%%%%%%%%%%%%%%%%%%%%%%%
%%%%%%%%%%%%%%%%%%%%%%%%%%%%%%%%%%%%%%%%%%%%%%%%%%%%%%%%%%%%
%%%%%%%%%%%%%%%%%%%%%%%%%%%%%%%%%%%%%%%%%%%%%%%%%%%%%%%%%%%%
%%%%%%%%%%%%%%%%%%%%%%%%%%%%%%%%%%%%%%%%%%%%%%%%%%%%%%%%%%%%

\section{N\lowercase{umerical implementation of the mean-field model}}
\label{appendix:numerical_implementation}

The mean-field model we studied is formulated through the disordered Hamiltonian $H$ defined in Eq.~\eqref{eq:Hamiltonian}. For all the numerical results reported in this work, samples were prepared by performing energy minimization in the absence of an applied force, $f(t)\!=\!0$, starting from random oscillator configurations where each $x_i$ is drawn from a uniform distribution over $[-\epsilon/2,\epsilon/2]$ with $\epsilon\!=\!10^{-2}$. We verified that the value of $\epsilon$ does not affect the results. We prepared $M\!=\!2000$ independent realizations of quenched disorder at $N\!=\!8192$ for the results presented in Figs.~\ref{fig_monopole_short:Fig2}b,\ref{fig_monopole_short:Fig3}--\ref{fig_monopole_short:Fig8},\ref{fig_monopole_short:FigS3}--\ref{fig_monopole_short:FigS5}, and $M\!=\!\{16000,8000,4000,2000,1000,500\}$ disorder realizations at $N\!=\!\{1024,2048,4096,8192,16000,32000\}$ for those presented in Figs.~\ref{fig_monopole_short:Fig2}a,\ref{fig_monopole_short:FigS1}--\ref{fig_monopole_short:FigS2}.

\section{F\lowercase{orce monopole responses in the spherical 3-spin model}}
\label{appendix:3spin_model}

In Sect.~\ref{sec:identity}, it was stated that a large class of mean-field models feature ${\cal D}_{\!_{\rm G}\!}(\omega)\!\sim\!\omega^2$ in the $\omega\!\to\!0$ limit, which implies an ill-defined monopole response, to be explicitly demonstrated for the spherical 3-spin model. Here, we support this statement. The Hamiltonian of the spherical 3-spin model reads~\cite{Derrida1981-nk,Crisanti1992-xn,Castellani2005-pd}
\begin{equation}
    H^{\text{(3-sp)}}(\bm s)\equiv\sum_{i<j<k}J_{ijk}s_is_js_k\,,\qquad\sum_is_i^2=N\,,
\label{eq:3spin_Hamiltonian}
\end{equation}
where $\bm s\!\equiv\!(s_1,...,s_N)$ lies on a sphere of radius $\sqrt{N}$, and the couplings $J_{ijk}$ are independent Gaussian random variables with a zero mean and variance $J^2/N^2$.

%%%%%%%%%%%%%%%%%%%%%%%%%%%%%%%%%%%%%%%%%%%%%%%%%%%%%%%%%%%%
\begin{figure}[t!]
    \centering
    \includegraphics[width=0.95\columnwidth]{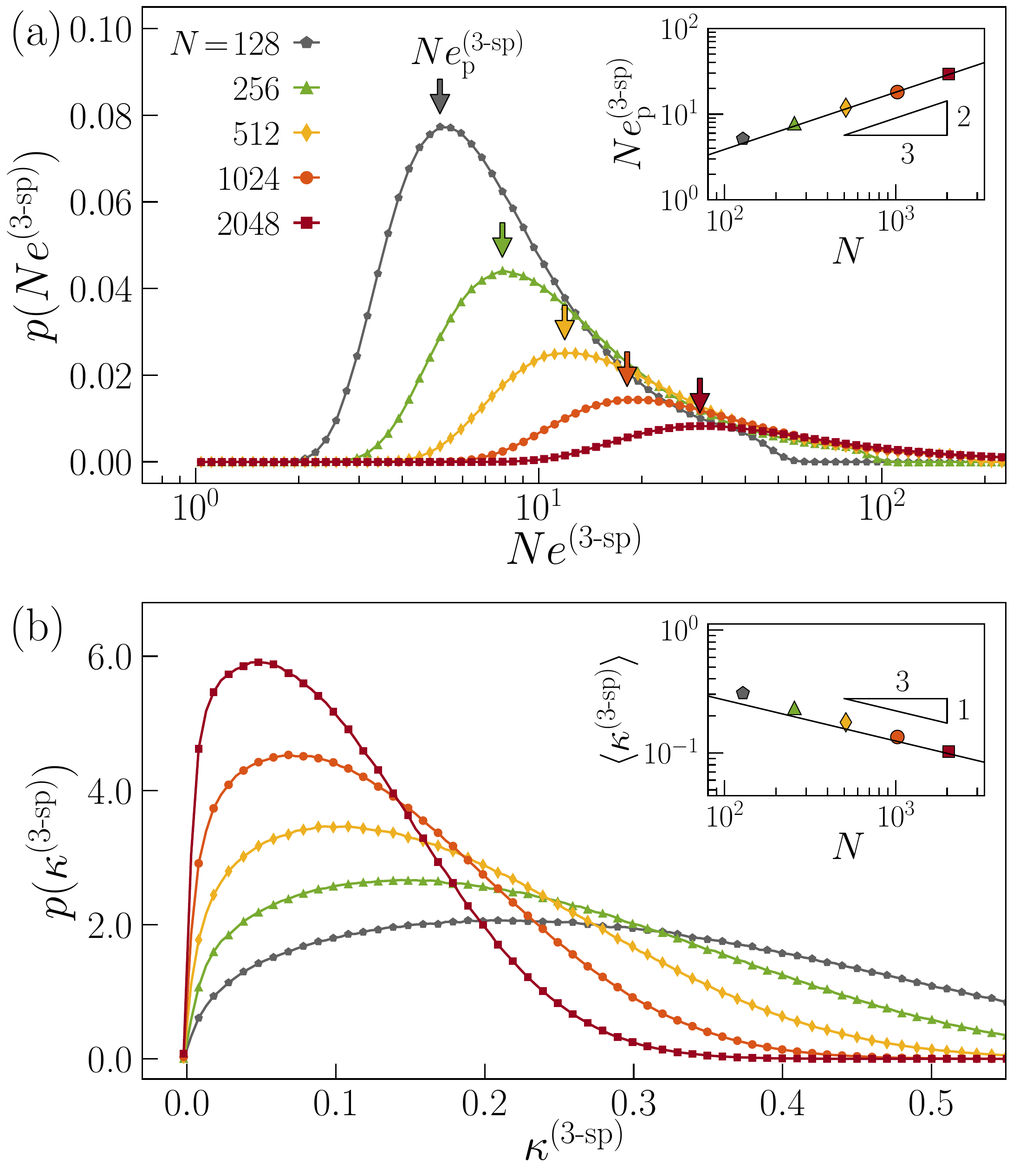}
    \caption{(a) The distribution $p(Ne^{\text{(3-sp)}})$ of the participation number of the monopole response in the spherical 3-spin model. Herein plotted for $J\!=\!1$ and various system sizes $N$ (see legend). The maximum of $p(Ne^{\text{(3-sp)}})$ is marked on each curve (see downward-pointing arrows) and denoted as $Ne_{\rm p}^{\text{(3-sp)}}$. (inset) The peak participation number $Ne_{\rm p}^{\text{(3-sp)}}$ vs. system size $N$. The superimposed solid line indicates the analytic prediction $Ne_{\rm p}^{\text{(3-sp)}}\!\sim\!N^{2/3}$. (b) The distribution $p(\kappa^{\text{(3-sp)}})$ of monopole stiffnesses in the spherical 3-spin model, plotted for the same ensemble as (a). (inset) The mean monopole stiffness $\langle\kappa^{\text{(3-sp)}}\rangle$ vs. system size $N$. The superimposed solid line indicates the analytic prediction $\langle\kappa^{\text{(3-sp)}}\rangle\!\sim\!N^{-1/3}$. See text for definitions and discussion.}
    \label{fig_monopole_short:FigS6}
\end{figure}
%%%%%%%%%%%%%%%%%%%%%%%%%%%%%%%%%%%%%%%%%%%%%%%%%%%%%%%%%%%%

The spherical constraint implies that  displacements are tangent to the sphere. Hence, we introduce the tangent-space projector $\bm P\!\equiv\!\bm I\!-\!N^{-1}\bm s\otimes\bm s$. The curvature relevant to linear response is then described by the tangent-space Hessian
\begin{equation}
    \bm{\mathcal{M}}^{\parallel} = \bm P\cdot\left[\frac{\pa^2 H^{\text{(3-sp)}}}{\pa \bm s\pa\bm s}-\left(\frac{\bm s}{N}\cdot\frac{\pa H^{\text{(3-sp)}}}{\pa \bm s}\right)\bm I\right]\cdot\bm P\,,
\end{equation}
where the superscript $\parallel$ denotes projection onto the tangent space. When represented in the full $N$-dimensional space, $\bm{\mathcal{M}}^\parallel$ has a zero mode along the radial direction $\bm s$. Its inverse is therefore understood as the pseudoinverse
\begin{equation}
(\bm{\mathcal{M}}^{\parallel})^{-1}\equiv\sum_{\lambda_\alpha>0}\lambda_\alpha^{-1}\bm\psi^{\parallel}_\alpha\otimes\bm\psi^{\parallel}_\alpha\,,
\end{equation}
where $\bm\psi^{\parallel}_\alpha$ and $\lambda_\alpha$ are the normalized tangent eigenmodes and corresponding eigenvalues, respectively.

We then define the local monopole response analogously to Sect.~\ref{sec:force_monople}. Projecting the Cartesian monopole $\bm f_i$ onto the tangent space gives $\bm f_i^\parallel\!\equiv\!\bm P\cdot\bm f_i$, and the corresponding response is $\bm z_i^\parallel\!\equiv\!(\bm{\mathcal{M}}^\parallel)^{-1}\cdot\bm f_i^\parallel$. Its effective stiffness is
\begin{equation}
    \kappa_i^{\text{(3-sp)}}\!\equiv\!\frac{\bm z_i^{\parallel}\cdot\bm f^{\parallel}_i}{\bm z_i^{\parallel}\cdot\bm z_i^{\parallel}} = \frac{\bm f^{\parallel}_i\cdot(\bm{\mathcal{M}}^{\parallel})^{-1}\cdot\bm f^{\parallel}_i}{\bm f^{\parallel}_i\cdot(\bm{\mathcal{M}}^{\parallel})^{-2}\cdot\bm f^{\parallel}_i}\,.
\label{eq:3spin_kappa}
\end{equation}

At the local minima of $H^{\text{(3-sp)}}$ sampled by gradient-based numerical minimization, the Hessian $\bm{\mathcal{M}}^{\parallel}$ is described by a shifted GOE matrix, where its VDoS is pseudogapped and follows ${\cal D}_{\!_{\rm G}\!}(\omega)\!\sim\!\omega^2$, and its eigenvectors $\bm \psi^{\parallel}_\alpha$ are uniformly distributed over a unit sphere in the tangent space according to the Haar measure~\cite{Auffinger2013-bm}. Consequently, the monopole stiffness $\kappa_i^{\text{(3-sp)}}$ is evaluated as
\begin{equation}
    \kappa_i^{\text{(3-sp)}}\sim\frac{\sum_{\lambda_\alpha>0} \lambda_\alpha^{-1}}{\sum_{\lambda_\alpha>0} \lambda_\alpha^{-2}}=\frac{\chi^{\text{(3-sp)}}}{\chi^{\text{(3-sp)}}_{_{\rm SG}}}\sim N^{-1/3}\,,
\label{eq:3spin_kappa_prediction}
\end{equation}
while the participation number of the response $Ne_i^{\text{(3-sp)}}\!\equiv\!(\sum_jz_{i,j}^2)^2/\sum_jz_{i,j}^4$ is evaluated as
\begin{equation}
    Ne_i^{\text{(3-sp)}}\sim\,\frac{(\chi_{_{\rm SG}}^{\text{(3-sp)}})^2}{(\chi^{\text{(3-sp)}})^4}\sim N^{2/3}\,.
\label{eq:3spin_Ne_prediction}
\end{equation}
The $N$-dependencies follow from the extreme value estimate $N^{-1}\!\sim\!\int_0^{\langle\omega_{\rm min}\rangle}{\cal D}_{\!_{\rm G}\!}(\omega)\,d\omega$ with ${\cal D}_{\!_{\rm G}\!}(\omega)\!\sim\!\omega^2$.

To test these predictions numerically, we draw configurations uniformly from the sphere of radius $\sqrt{N}$ and minimize the Hamiltonian in Eq.~\eqref{eq:3spin_Hamiltonian}. For each sampled minimum, we then evaluate $Ne_i^{\text{(3-sp)}}$ and $\kappa_i^{\text{(3-sp)}}$ as defined above. Figure~\ref{fig_monopole_short:FigS6} shows the resulting distributions at $J\!=\!1$ for system sizes $N\!=\!\{128,256,512,1024,2048\}$, averaged over $M\!=\!\{64000,32000,16000,8000,4000\}$ disorder realizations. With increasing $N$, the participation number increases, and the monopole stiffness shifts to smaller values. The peak participation number $Ne_{\rm p}^{\text{(3-sp)}}$ and the mean monopole stiffness $\langle\kappa^{\text{(3-sp)}}\rangle$, shown in the insets, follow the predicted $N$-scalings in Eqs.~\eqref{eq:3spin_kappa_prediction} and~\eqref{eq:3spin_Ne_prediction}. Overall, these results show that ${\cal D}_{\!_{\rm G}\!}(\omega)\!\sim\!\omega^2$ spectra found in many mean-field models do not support a well-defined local linear response and the associated micromechanical description discussed in this work.

%\clearpage

%\bibliography{Monopole_response_short}


\end{document}